\documentclass[manuscript]{aastex}

\usepackage{amsbsy,amssymb,amsmath,bm}
\usepackage{graphicx,subfigure}
\usepackage{natbib}
\usepackage{color}
\usepackage{multirow}

\begin{document}

\title{Wave reflection and transmission at interface of convective and stably stratified regions in a rotating star or planet}
\author{Xing Wei}
\affil{Department of Astronomy, Beijing Normal University, Beijing, China}
\email{xingwei@bnu.edu.cn}

\begin{abstract}
We use a simplified model to study wave reflection and transmission at interface of convective region and stably stratified region (e.g. radiative zone in star or stratification layer in gaseous planet). Inertial wave in convective region and gravito-inertial wave in stably stratified region are considered. We begin with polar area and then extend to any latitude. Six cases are discussed (see Table \ref{tab1}), and in Case 2 both waves co-exist in both regions. Four configurations are further discussed for Case 2. The angles and energy ratios of wave reflection and transmission are calculated. It is found that wave propagation and transmission depend on the ratio of buoyancy frequency to rotational frequency. In a rapidly rotating star or planet wave propagates across interface and most of energy of incident wave is transmitted to the other region, but in a slowly rotating star or planet wave transmission is inhibited.
\end{abstract}
\maketitle

\section{Motivation}
The interior of a star or planet has layered structure. In a solar-type star convective zone sits on radiative zone while in a massive star radiative zone sits on convective zone. Radiative zone can be treated as a stably stratified region with the diffusion limit. In some gaseous planets such as Saturn a stably stratified layer sits on convective region \citep{fuller} and this layer is believed to filter out the non-axisymmetric components of magnetic field \citep{cao}. In a rotating star or planet, inertial wave (r mode) can be induced by Coriolis force in convective region \citep{ogilvie04,wu,goodman09,ogilvie14}. In a non-rotating star or planet, gravity wave (g mode) can be induced by density stratification in stably stratified region \citep{zahn,goldreich,goodman98}. In a rotating star or planet, gravito-inertial wave can be induced in stably stratified region due to the combined effect of rotation and density stratification \citep{rieutord}.

An interesting problem is how wave reflects and transmits at interface of two regions. In the solar interior, it is believed that g mode in radiative region propagating outward cannot penetrate deep into convective region such that it is difficult to find g mode on the solar surface \citep{science}. However, in some other stars or planets, can wave in one region transmit to the other region? If stratified region is replaced with rigid region, then inertial wave in convective region totally reflects at interface \citep{goodman09}. However, for ``soft'' stratified region, how much wave energy can be transmitted and how much can be reflected? In this paper I will try to answer this question with a simplified model qualitatively and quantitatively.

\section{Model}
A simplified local geometry is studied in this paper. Convective region and stably stratified region (e.g. radiative zone in stellar interior or stratification layer in gaseous planet) are on two sides of an infinite plane which represents interface of the two regions. The rotational axis is normal to the plane, which simplifies the calculations. This is only valid near the vicinity of polar area and we will discuss in Section \S\ref{sec:latitude} the more general situation that rotational axis is not necessarily normal to interface. Sound wave is too fast to interact with inertial or internal gravity wave, and therefore we consider an incompressible fluid to filter out sound wave. In this local geometry, the global spherical curvature is not considered. However, we will see in the next texts that some interesting physics can be found in this simplified geometry.

In convective region, the linear perturbation equations of momentum and mass conservation in a rotating frame read
\begin{equation}\label{conv1}
\left\{
\begin{aligned}
\frac{\partial u_x'}{\partial t}&=-\frac{1}{\rho}\frac{\partial p'}{\partial x}+2\Omega u_y' \\
\frac{\partial u_y'}{\partial t}&=-\frac{1}{\rho}\frac{\partial p'}{\partial y}-2\Omega u_x' \\
\frac{\partial u_z'}{\partial t}&=-\frac{1}{\rho}\frac{\partial p'}{\partial z} \\
\frac{\partial u_x'}{\partial x}&+\frac{\partial u_y'}{\partial y}+\frac{\partial u_z'}{\partial z}=0
\end{aligned}
\right.
\end{equation}
The variables are written in their conventional manner and prime denotes Eulerian perturbation. Equations \eqref{conv1} describe the inertial wave caused by Coriolis force. In stably stratified region, the linear perturbation equations read
\begin{equation}\label{stra1}
\left\{
\begin{aligned}
\frac{\partial u_x'}{\partial t}&=-\frac{1}{\rho}\frac{\partial p'}{\partial x}+2\Omega u_y' \\
\frac{\partial u_y'}{\partial t}&=-\frac{1}{\rho}\frac{\partial p'}{\partial y}-2\Omega u_x' \\
\frac{\partial u_z'}{\partial t}&=-\frac{1}{\rho}\frac{\partial p'}{\partial z}-\frac{\rho'}{\rho}g \\
\frac{\partial \rho'}{\partial t}&+\beta u_z'=0 \\
\frac{\partial u_x'}{\partial x}&+\frac{\partial u_y'}{\partial y}+\frac{\partial u_z'}{\partial z}=0
\end{aligned}
\right.
\end{equation}
Equations \eqref{stra1} describe gravito-inertial wave. Compared to \eqref{conv1}, a gravity term appears in $z$ direction and an equation of density stratification is added. The parameter 
\begin{equation}\label{beta}
\beta=\frac{d\rho}{dz}<0
\end{equation}
measures stratification and it is assumed to be constant in the local geometry, i.e. the WKB approximation due to the short wavelength of perturbations compared to the length scale of stratification. That is, the density is continuous across interface but the density gradient is not. Here the Boussinesq approximation is used, i.e. in equation of motion density is constant and density variation is considered only in the gravity term, and density stratification is only considered with $\beta$. This approximation is valid only for small variation of density. Usually in stars or gaseous planets this approximation cannot hold, and fully compressible fluid or anelastic approximation should be considered. However, in a very small local region near interface Boussinesq approximation still works.

We apply normal mode analysis to the linear perturbation equations. All the perturbations are expressed as a Fourier plane wave with its complex amplitude dependent on $z$ coordinate, e.g.
\begin{equation}\label{normal}
u_z'=\Re\{\hat u_z(z) \exp\left[i(k_x x+k_y y-\omega t)\right]\},
\end{equation}
where hat denotes the complex amplitude and $\Re$ denotes the real part of a complex variable. Then the two sets of amplitude equations are derived as follows,
\begin{equation}\label{conv2}
\left\{
\begin{aligned}
-i\omega \hat u_x&=-\frac{1}{\rho}ik_x \hat p+2\Omega \hat u_y \\
-i\omega \hat u_y&=-\frac{1}{\rho}ik_y \hat p-2\Omega \hat u_x \\
-i\omega \hat u_z&=-\frac{1}{\rho}D\hat p \\
ik_x \hat u_x&+ik_y \hat u_y+D\hat u_z=0
\end{aligned}
\right.
\end{equation}
in convective region, and
\begin{equation}\label{stra2}
\left\{
\begin{aligned}
-i\omega \hat u_x&=-\frac{1}{\rho}ik_x \hat p+2\Omega \hat u_y \\
-i\omega \hat u_y&=-\frac{1}{\rho}ik_y \hat p-2\Omega \hat u_x \\
-i\omega \hat u_z&=-\frac{1}{\rho}D\hat p-\frac{\hat\rho}{\rho}g \\
-i\omega \hat\rho&+\beta \hat u_z=0 \\
ik_x \hat u_x&+ik_y \hat u_y+D\hat u_z=0
\end{aligned}
\right.
\end{equation}
in stably stratified region. In \eqref{conv2} and \eqref{stra2} $D$ denotes $d/dz$ and $\Re$ is omitted because of the linear property of equations.

Either \eqref{conv2} or \eqref{stra2} can be further reduced to a single equation with $\hat u_z$ being variable, i.e.
\begin{equation}\label{conv3}
D^2 \hat u_z+\frac{\omega^2 k^2}{4\Omega^2-\omega^2} \hat u_z=0
\end{equation}
in convective region, where $k^2 = k_x^2 + k_y^2$, and
\begin{equation}\label{stra3}
D^2 \hat u_z+\frac{(\omega^2-N^2) k^2}{4\Omega^2-\omega^2} \hat u_z=0
\end{equation}
in stably stratified region, where
\begin{equation}
N^2=-\frac{\beta g}{\rho}
\end{equation}
is the square of buoyancy frequency. In the derivation of \eqref{stra3} the first-order term $D\hat u_z$ is neglected under the Boussinesq approximation in which density $\rho$ and density stratification $\beta$ are both constants. When $N=0$ Equation \eqref{stra3} reduces to \eqref{conv3}. Equations \eqref{conv3} and \eqref{stra3} are what we will discuss in the next texts. With the solution of $\hat u_z$ the two horizontal components are derived as follows,
\begin{equation}\label{uxuy1}
\begin{aligned}
\hat u_x&=-\frac{\omega k_x+2i\Omega k_y}{i\omega k^2}D\hat u_z, \\
\hat u_y&=\frac{2i\Omega k_x-\omega k_y}{i\omega k^2}D\hat u_z.
\end{aligned}
\end{equation}
The two boundary conditions are imposed. One is that vertical velocity is continuous across interface, i.e.
\begin{equation}\label{b1}
\hat u_z(0^+)=\hat u_z(0^-).
\end{equation}
The other is that the Lagrangian perturbation of pressure, which is identical to Eulerian perturbation of pressure, is continuous across interface. According to the $x$ and $y$ components of momentum conservation equation and the mass conservation equation, this condition can be translated to
\begin{equation}\label{b2}
D\hat u_z(0^+)=D\hat u_z(0^-).
\end{equation}

\section{Six cases}
According to \eqref{conv3}, to support wave motion in convective region it is required that $4\Omega^2-\omega^2>0$. If $4\Omega^2-\omega^2<0$ then wave amplitude exponentially decays away from interface. Similarly, to support wave motion in stably stratified region it is required that $(\omega^2-N^2)(4\Omega^2-\omega^2)>0$. Consequently, if waves exist in both regions then it is required that 
\begin{equation}\label{cond}
N^2<\omega^2<4\Omega^2.
\end{equation}
Condition \eqref{cond} implies that $N^2<4\Omega^2$, i.e. rotation wins out stratification. If stratification is so strong to win out rotation, i.e. $4\Omega^2<N^2$, then wave could exist only in one region but cannot exist in both regions. 

All the six cases are summarized in Table \ref{tab1}. In Cases 1 and 4 wave does not exist. In Case 2 waves exist in both regions. In Case 3, 5 or 6 only one wave exists in one region and in the other region wave amplitude exponentially decays away from interface, i.e. the so-called evanescent region. For example, Cases 5 and 6 take place in the solar interior.
\begin{table}
\centering
\begin{tabular}{|c|c|c|c|}
\hline
& & inertial wave & gravito-inertial wave \\
\hline
\multirow{3}{*}{$N^2<4\Omega^2$} & Case 1: $N^2<4\Omega^2<\omega^2$ & no & no \\
\cline{2-4}
& Case 2: $N^2<\omega^2<4\Omega^2$ & yes & yes \\
\cline{2-4}
& Case 3: $\omega^2<N^2<4\Omega^2$ & yes & no \\
\hline
\multirow{3}{*}{$4\Omega^2<N^2$} & Case 4: $4\Omega^2<N^2<\omega^2$ & no & no \\
\cline{2-4}
& Case 5: $4\Omega^2<\omega^2<N^2$ & no & yes \\
\cline{2-4}
& Case 6: $\omega^2<4\Omega^2<N^2$ & yes & no \\
\hline
\end{tabular}
\caption{Table of all the six cases for frequencies}\label{tab1}
\end{table}

\section{Case 2}
We now focus on Case 2, i.e. waves exist in both regions. The general solution in convective region is 
\begin{equation}
u_z'=\Re\left\{a_1\exp\left[i(k_x x+k_y y+qz-\omega t)\right]+a_2\exp\left[i(k_x x+k_y y-qz-\omega t)\right]\right\}
\end{equation}
where 
\begin{equation}\label{q}
q=\frac{\omega k}{\sqrt{4\Omega^2-\omega^2}},
\end{equation}
and the general solution in stably stratified region is 
\begin{equation}
u_z'=\Re\left\{b_1\exp\left[i(k_x x+k_y y+sz-\omega t)\right]+b_2\exp\left[i(k_x x+k_y y-sz-\omega t)\right]\right\}
\end{equation}
where 
\begin{equation}\label{s}
s=\sqrt{\frac{\omega^2-N^2}{4\Omega^2-\omega^2}}k.
\end{equation}
Note that $s<q$. The coefficients $a_1$, $a_2$, $b_1$ and $b_2$ will be determined by the boundary conditions \eqref{b1} and \eqref{b2} \citep{brekhovskikh}.

\begin{figure}
\centering
\includegraphics[scale=0.5]{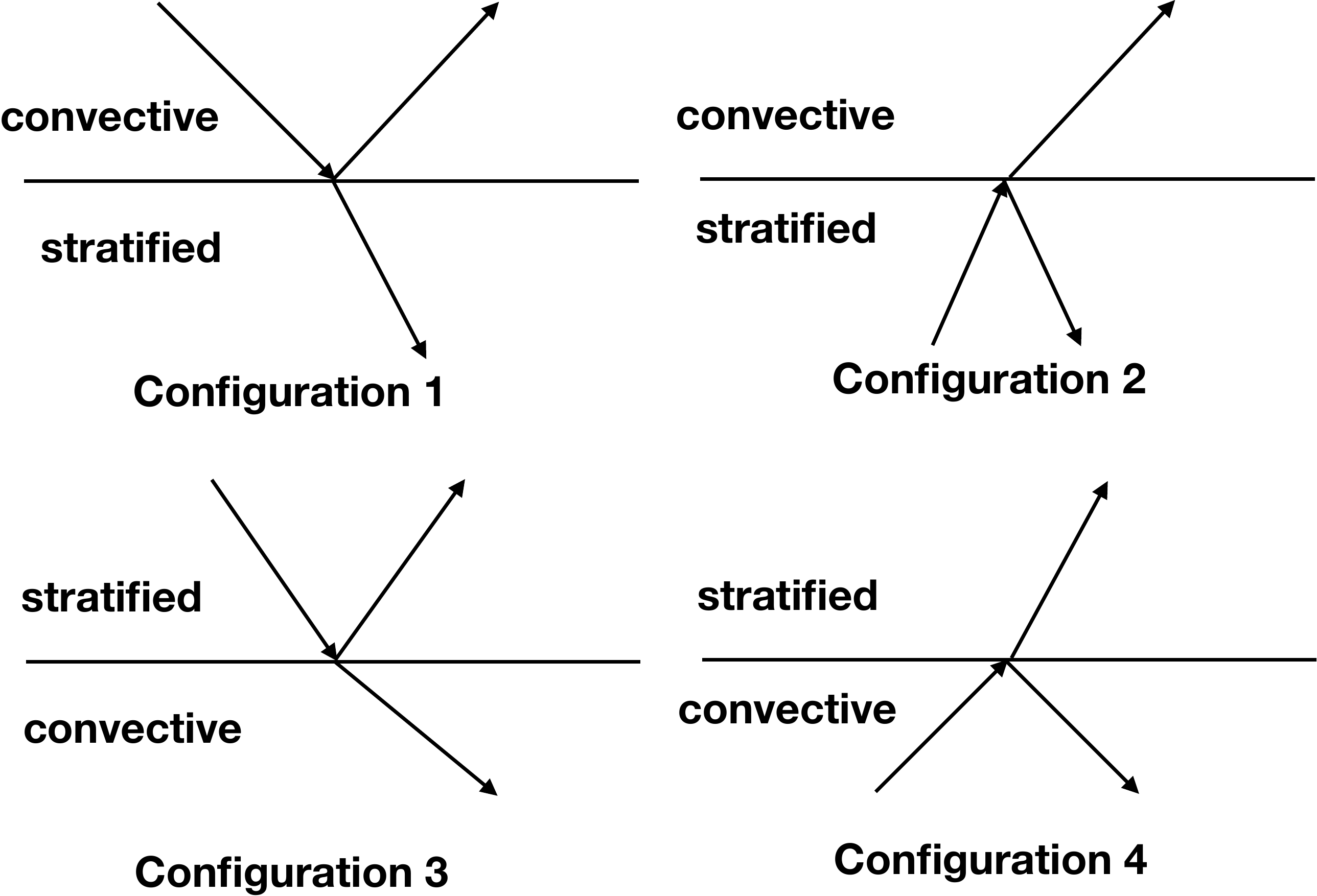}
\caption{Four configurations. 1 and 4 are categorized together, while 2 and 3 are categeorized together.}\label{fig1}
\end{figure}

We then discuss the following four configurations as shown in Figure \ref{fig1}:
\begin{itemize}
\item Configuration 1: convective region on top of stratified region and wave from convective region to stratified region;
\item Configuration 2: convective region on top of stratified region and wave from stratified region to convective region;
\item Configuration 3: stratified region on top of convective region and wave from stratified region to convective region;
\item Configuration 4: stratified region on top of convective region and wave from convective region to stratified region;
\end{itemize}
Configurations 1 and 2 are applied to solar-type stars, and Configurations 3 and 4 are applied to massive stars and gaseous planets such as Saturn.

\subsection{Configuration 1}
When incident wave comes from convective region, its phase velocity is downward, the phase velocity of reflected wave in convective region is upward, and the phase velocity of transmitted wave in stably stratified region is downward. Therefore, $b_1=0$ and 
\begin{equation}
\left\{
\begin{aligned}
\mbox{incident wave: }&u_z'=\Re\left\{a_2\exp\left[i(k_x x+k_y y-qz-\omega t)\right]\right\} \\
\mbox{reflected wave: }&u_z'=\Re\left\{a_1\exp\left[i(k_x x+k_y y+qz-\omega t)\right]\right\} \\
\mbox{transmitted wave: }&u_z'=\Re\left\{b_2\exp\left[i(k_x x+k_y y-sz-\omega t)\right]\right\}
\end{aligned}
\right.
\end{equation}
Then the boundary conditions \eqref{b1} and \eqref{b2} yield the relationship
\begin{equation}\label{relation1}
a_1=\frac{q-s}{q+s}a_2, \; b_2=\frac{2q}{q+s}a_2.
\end{equation}

The three wave solutions can be then obtained as follows, 
\begin{equation}
\mbox{incident wave:}
\left\{
\begin{aligned}
u_x'&=\Re\left\{q\frac{\omega k_x+2i\Omega k_y}{\omega k^2}a_2\exp\left[i(k_x x+k_y y-qz-\omega t)\right]\right\} \\
u_y'&=\Re\left\{-q\frac{2i\Omega k_x-\omega k_y}{\omega k^2}a_2\exp\left[i(k_x x+k_y y-qz-\omega t)\right]\right\} \\
u_z'&=\Re\left\{a_2\exp\left[i(k_x x+k_y y-qz-\omega t)\right]\right\}
\end{aligned}
\right.
\end{equation}
\begin{equation}
\mbox{reflected wave:}
\left\{
\begin{aligned}
u_x'&=\Re\left\{-q\frac{\omega k_x+2i\Omega k_y}{\omega k^2}a_1\exp\left[i(k_x x+k_y y+qz-\omega t)\right]\right\} \\
u_y'&=\Re\left\{q\frac{2i\Omega k_x-\omega k_y}{\omega k^2}a_1\exp\left[i(k_x x+k_y y+qz-\omega t)\right]\right\} \\
u_z'&=\Re\left\{a_1\exp\left[i(k_x x+k_y y+qz-\omega t)\right]\right\}
\end{aligned}
\right.
\end{equation}
\begin{equation}
\mbox{transmitted wave:}
\left\{
\begin{aligned}
u_x'&=\Re\left\{s\frac{\omega k_x+2i\Omega k_y}{\omega k^2}b_2\exp\left[i(k_x x+k_y y-sz-\omega t)\right]\right\} \\
u_y'&=\Re\left\{-s\frac{2i\Omega k_x-\omega k_y}{\omega k^2}b_2\exp\left[i(k_x x+k_y y-sz-\omega t)\right]\right\} \\
u_z'&=\Re\left\{b_2\exp\left[i(k_x x+k_y y-sz-\omega t)\right]\right\}
\end{aligned}
\right.
\end{equation}
where $k=\sqrt{k_x^2+k_y^2}$.

By the three wave solutions, we can find more information. Firstly we calculate the angles of three waves to $z$ axis. Clearly, incident and reflected waves have the same angle but transmitted wave has a different one. The reflected and transmitted angles are, respectively,
\begin{equation}
\arctan\left({\frac{q}{k}\sqrt{1+\frac{4\Omega^2}{\omega^2}}}\right), \;\arctan\left({\frac{s}{k}\sqrt{1+\frac{4\Omega^2}{\omega^2}}}\right).
\end{equation}
It shows that transmitted wave has a smaller angle because $s<q$. The time-averaged energies 
\begin{equation}\label{energy}
\frac{\omega}{2\pi}\int_0^{2\pi/\omega}(u_x'^2+u_y'^2+u_z'^2)dt=\frac{1}{2}\left(|\hat u_x|^2+|\hat u_y|^2+|\hat u_z|^2\right)
\end{equation}
of three waves are, respectively,
\begin{equation}
a_2^2\frac{4\Omega^2}{4\Omega^2-\omega^2}, \; a_1^2\frac{4\Omega^2}{4\Omega^2-\omega^2}, \; b_2^2\frac{4\Omega^2-N^2/2-2\Omega^2 N^2/\omega^2}{4\Omega^2-\omega^2}
\end{equation}
for incident wave, reflected wave and transmitted wave. We then calculate the reflection ratio, i.e. the ratio of reflected wave energy to incident wave energy, 
\begin{equation}\label{ratio1}
\left(\frac{a_1}{a_2}\right)^2=\left(\frac{q-s}{q+s}\right)^2=\left(\frac{1-\sqrt{1-N^2/\omega^2}}{1+\sqrt{1-N^2/\omega^2}}\right)^2.
\end{equation}
It is very interesting that the energy ratio of reflected wave in convective region is independent of rotation rate but is influenced by the buoyancy frequency in stably stratified region. When $N\ll\omega<2\Omega$ the reflection ratio is approximately equal to $N^4/16\omega^4\approx 0$. The transmission ratio can be also calculated to be the ratio of transmitted wave energy to incident wave energy,
\begin{equation}\label{ratio2}
\left(\frac{b_2}{a_2}\right)^2\frac{4\Omega^2-N^2/2-2\Omega^2 N^2/\omega^2}{4\Omega^2}=\frac{4}{\left(1+\sqrt{1-N^2/\omega^2}\right)^2}\left(1-N^2/8\Omega^2-N^2/2\omega^2\right).
\end{equation}
When $N\ll\omega<2\Omega$ the transmission ratio is approximately equal to $1$. Therefore, we obtain an important result, i.e. when stratification is very weak or rotation is very strong ($N\ll\omega<2\Omega$), inertial wave from convective region can be almost transmitted to stably stratified region and changes to gravito-inertial wave.

\subsection{Configuration 2}
In this configuration, $a_2=0$ and 
\begin{equation}
\left\{
\begin{aligned}
\mbox{incident wave: }&u_z'=\Re\left\{b_1\exp\left[i(k_x x+k_y y+sz-\omega t)\right]\right\} \\
\mbox{reflected wave: }&u_z'=\Re\left\{b_2\exp\left[i(k_x x+k_y y-sz-\omega t)\right]\right\} \\
\mbox{transmitted wave: }&u_z'=\Re\left\{a_1\exp\left[i(k_x x+k_y y+qz-\omega t)\right]\right\}
\end{aligned}
\right.
\end{equation}
The relation of coefficients is
\begin{equation}\label{relation2}
b_2=-\frac{q-s}{q+s}b_1, \; a_1=\frac{2s}{q+s}b_1.
\end{equation}

We do not show readers the three wave solutions but directly give the wave angles to $z$ axis and the time-averaged wave energies. The reflected and transmitted angles to $z$ axis are, respectively,
\begin{equation}
\arctan\left({\frac{s}{k}\sqrt{1+\frac{4\Omega^2}{\omega^2}}}\right), \;\arctan\left({\frac{q}{k}\sqrt{1+\frac{4\Omega^2}{\omega^2}}}\right).
\end{equation}
In this configuration the transmitted angle is larger. The time-averaged wave energies are, respectively,
\begin{equation}
b_1^2\frac{4\Omega^2-N^2/2-2\Omega^2 N^2/\omega^2}{4\Omega^2-\omega^2}, \; b_2^2\frac{4\Omega^2-N^2/2-2\Omega^2 N^2/\omega^2}{4\Omega^2-\omega^2}, \; a_1^2\frac{4\Omega^2}{4\Omega^2-\omega^2}
\end{equation}
for incident wave, reflected wave and transmitted wave. The reflection ratio can be readily calculated, 
\begin{equation}\label{ratio3}
\left(\frac{b_2}{b_1}\right)^2=\left(\frac{q-s}{q+s}\right)^2=\left(\frac{1-\sqrt{1-N^2/\omega^2}}{1+\sqrt{1-N^2/\omega^2}}\right)^2.
\end{equation}
When $N\ll\omega<2\Omega$ the reflection ratio is approximately equal to $N^4/16\omega^4\approx 0$. The transmission ratio is
\begin{equation}\label{ratio4}
\left(\frac{a_1}{b_1}\right)^2\frac{4\Omega^2}{4\Omega^2-N^2/2-2\Omega^2 N^2/\omega^2}=4\left(\frac{\sqrt{1-N^2/\omega^2}}{1+\sqrt{1-N^2/\omega^2}}\right)^2\frac{1}{1-N^2/8\Omega^2-N^2/2\omega^2}.
\end{equation}
When $N\ll\omega<2\Omega$ the transmission ratio is approximately equal to $1$, i.e. gravito-inertial wave from stably stratified region is almost transmitted to convective region and changes to inertial wave.

\subsection{Configuration 3}
As before, we show readers the results. $a_1=0$ and the three waves are
\begin{equation}
\left\{
\begin{aligned}
\mbox{incident wave: }&u_z'=\Re\left\{b_2\exp\left[i(k_x x+k_y y-sz-\omega t)\right]\right\} \\
\mbox{reflected wave: }&u_z'=\Re\left\{b_1\exp\left[i(k_x x+k_y y+sz-\omega t)\right]\right\} \\
\mbox{transmitted wave: }&u_z'=\Re\left\{a_2\exp\left[i(k_x x+k_y y-qz-\omega t)\right]\right\}
\end{aligned}
\right.
\end{equation}
The boundary conditions yield
\begin{equation}
b_1=-\frac{q-s}{q+s}b_2, \; a_2=\frac{2s}{q+s}b_2.
\end{equation}
The reflected and transmitted angles as well as the reflection and transmission ratios are identical to those of Configuration 2. When $N\ll\omega<2\Omega$, gravito-inertial wave almost transmits into convective region and changes to inertial wave.

\subsection{Configuration 4}
In this configuration, $b_2=0$ and the three waves are
\begin{equation}
\left\{
\begin{aligned}
\mbox{incident wave: }&u_z'=\Re\left\{a_1\exp\left[i(k_x x+k_y y+qz-\omega t)\right]\right\} \\
\mbox{reflected wave: }&u_z'=\Re\left\{a_2\exp\left[i(k_x x+k_y y-qz-\omega t)\right]\right\} \\
\mbox{transmitted wave: }&u_z'=\Re\left\{b_1\exp\left[i(k_x x+k_y y+sz-\omega t)\right]\right\}
\end{aligned}
\right.
\end{equation}
The boundary conditions yield
\begin{equation}
a_2=\frac{q-s}{q+s}a_1, \; b_1=\frac{2q}{q+s}a_1.
\end{equation}
The three angles to $z$ axis are
\begin{equation}
\arctan\left({\frac{q}{k}\sqrt{1+\frac{4\Omega^2}{\omega^2}}}\right), \; \arctan\left({\frac{q}{k}\sqrt{1+\frac{4\Omega^2}{\omega^2}}}\right), \;\arctan\left({\frac{s}{k}\sqrt{1+\frac{4\Omega^2}{\omega^2}}}\right).
\end{equation}
The transmitted wave has a smaller angle. The three energies are
\begin{equation}
a_1^2\frac{4\Omega^2}{4\Omega^2-\omega^2}, \; a_2^2\frac{4\Omega^2}{4\Omega^2-\omega^2}, \; b_1^2\frac{4\Omega^2-N^2/2-2\Omega^2 N^2/\omega^2}{4\Omega^2-\omega^2}.
\end{equation}
The angles and ratios are identical to those of Configuration 1.When $N\ll\omega<2\Omega$, inertial wave almost transmits into stratified region and changes to gravito-inertial wave.

\subsection{Summary}
We studied the six cases as listed in Tabel \ref{tab1}, and in Case 2 ($N^2<\omega^2<4\Omega^2$) both convective and stratified regions support wave motion. We then focused on the four configurations in Case 2 for solar-type and massive stars. It is found that 
\begin{itemize}
\item Configurations 1 and 4 are symmetric (wave propagation from convective region to stratified region) while Configurations 2 and 3 are symmetric (wave propagation from stratified region to convective region);
\item the wave angle in convective region is larger than that in stratified region;
\item in a rapidly rotating star or planet ($N\ll\Omega$) wave reflects very little and almost transmits to the other region.
\end{itemize}

\section{At any latitude}\label{sec:latitude}
As mentioned in Section Model, in our simplified model the rotational axis is normal to interface, which is valid only in the vicinity of polar area. We now discuss the situation at any latitude $\theta$. In the local plane model, we assign that $x$ directs east, $y$ directs north and $z$ directs radially outward. At latitude $\theta$, angular velocity in the local Cartesian coordinates is written as $(0,\Omega\cos\theta,\Omega\sin\theta)$. Then Coriolis force depends on $\theta$ and then the two sets of perturbation equations are 
\begin{equation}
\left\{
\begin{aligned}
\frac{\partial u_x'}{\partial t}&=-\frac{1}{\rho}\frac{\partial p'}{\partial x}+2\Omega (u_y'\sin\theta-u_z'\cos\theta) \\
\frac{\partial u_y'}{\partial t}&=-\frac{1}{\rho}\frac{\partial p'}{\partial y}-2\Omega u_x'\sin\theta \\
\frac{\partial u_z'}{\partial t}&=-\frac{1}{\rho}\frac{\partial p'}{\partial z}+2\Omega u_x'\cos\theta \\
\frac{\partial u_x'}{\partial x}&+\frac{\partial u_y'}{\partial y}+\frac{\partial u_z'}{\partial z}=0
\end{aligned}
\right.
\end{equation}
in convective region and
\begin{equation}
\left\{
\begin{aligned}
\frac{\partial u_x'}{\partial t}&=-\frac{1}{\rho}\frac{\partial p'}{\partial x}+2\Omega (u_y'\sin\theta-u_z'\cos\theta) \\
\frac{\partial u_y'}{\partial t}&=-\frac{1}{\rho}\frac{\partial p'}{\partial y}-2\Omega u_x'\sin\theta \\
\frac{\partial u_z'}{\partial t}&=-\frac{1}{\rho}\frac{\partial p'}{\partial z}+2\Omega u_x'\cos\theta-\frac{\rho'}{\rho}g \\
\frac{\partial \rho'}{\partial t}&+\beta u_z'=0 \\
\frac{\partial u_x'}{\partial x}&+\frac{\partial u_y'}{\partial y}+\frac{\partial u_z'}{\partial z}=0
\end{aligned}
\right.
\end{equation}
in stratified region. Next, following the above procedure we are led to
\begin{equation}\label{conv4}
D^2 \hat u_z+i\frac{8\Omega^2k_y\sin\theta\cos\theta}{4\Omega^2\sin^2\theta-\omega^2}D\hat u_z+\frac{\omega^2 k^2-4\Omega^2k_y^2\cos^2\theta}{4\Omega^2\sin^2\theta-\omega^2} \hat u_z=0
\end{equation}
in convective region and
\begin{equation}\label{stra4}
D^2 \hat u_z+i\frac{8\Omega^2k_y\sin\theta\cos\theta}{4\Omega^2\sin^2\theta-\omega^2}D\hat u_z+\frac{(\omega^2-N^2) k^2-4\Omega^2k_y^2\cos^2\theta}{4\Omega^2\sin^2\theta-\omega^2} \hat u_z=0
\end{equation}
in stratified region. When $\theta=90^\circ$ Equations \eqref{conv4} and \eqref{stra4} reduce to \eqref{conv3} and \eqref{stra3} respectively. 

In \eqref{conv4} and \eqref{stra4} $k_y$ appears, which suggests the importance of the direction of wave vector. Denote the angle of wave vector to the east (i.e. $x$ direction) on the local horizontal plane by $\alpha$, then 
\begin{equation}
k_x=k\cos\alpha, \; k_y=k\sin\alpha.
\end{equation} 
The condition that wave propagates in both regions (Case 2) is that \eqref{conv4} and \eqref{stra4} admit wave solution, which yields two inequalities,
\begin{equation}
\omega^2<\omega_0^2
\end{equation}
in convective region and 
\begin{equation}
\omega_1^2<\omega^2<\omega_2^2
\end{equation}
in stratified region, where 
\begin{align}
\omega_0^2&=4\Omega^2(\sin^2\theta+\cos^2\theta\sin^2\alpha), \\
\omega_{1,2}^2&=\frac{1}{2}\left[\omega_0^2+N^2\mp\sqrt{(\omega_0^2+N^2)^2-16\Omega^2N^2\sin^2\theta}\right].
\end{align}
Note that the frequency ranges $\omega_0$, $\omega_1$ and $\omega_2$ all depend on both $\theta$ and $\alpha$. More analysis shows that $\omega_1^2\leq\omega_0^2\leq\omega_2^2$. Therefore, the condition for the wave propagation across interface is 
\begin{equation}
|\omega_1|<|\omega|<|\omega_0|.
\end{equation}

Figure \ref{fig2} shows the contours of $(|\omega_0|-|\omega_1|)/2\Omega$ versus $\theta$ and $\alpha$ at $N/2\Omega=0.1$, 1 and 10. The darker region of a wider waveband $|\omega_0|-|\omega_1|$ implies more possibility for wave transmission. Fast rotation (small $N/2\Omega=0.1$) tends to support wave transmission anywhere, whereas slow rotation (large $N/2\Omega=10$) tends to support wave transmission only in the equatorial area. Moreover,  the wave with a longer longitudinal wavelength (larger $\alpha$) seems more possible to transmit.
\begin{figure}
\centering
\subfigure[]{\includegraphics[scale=0.28]{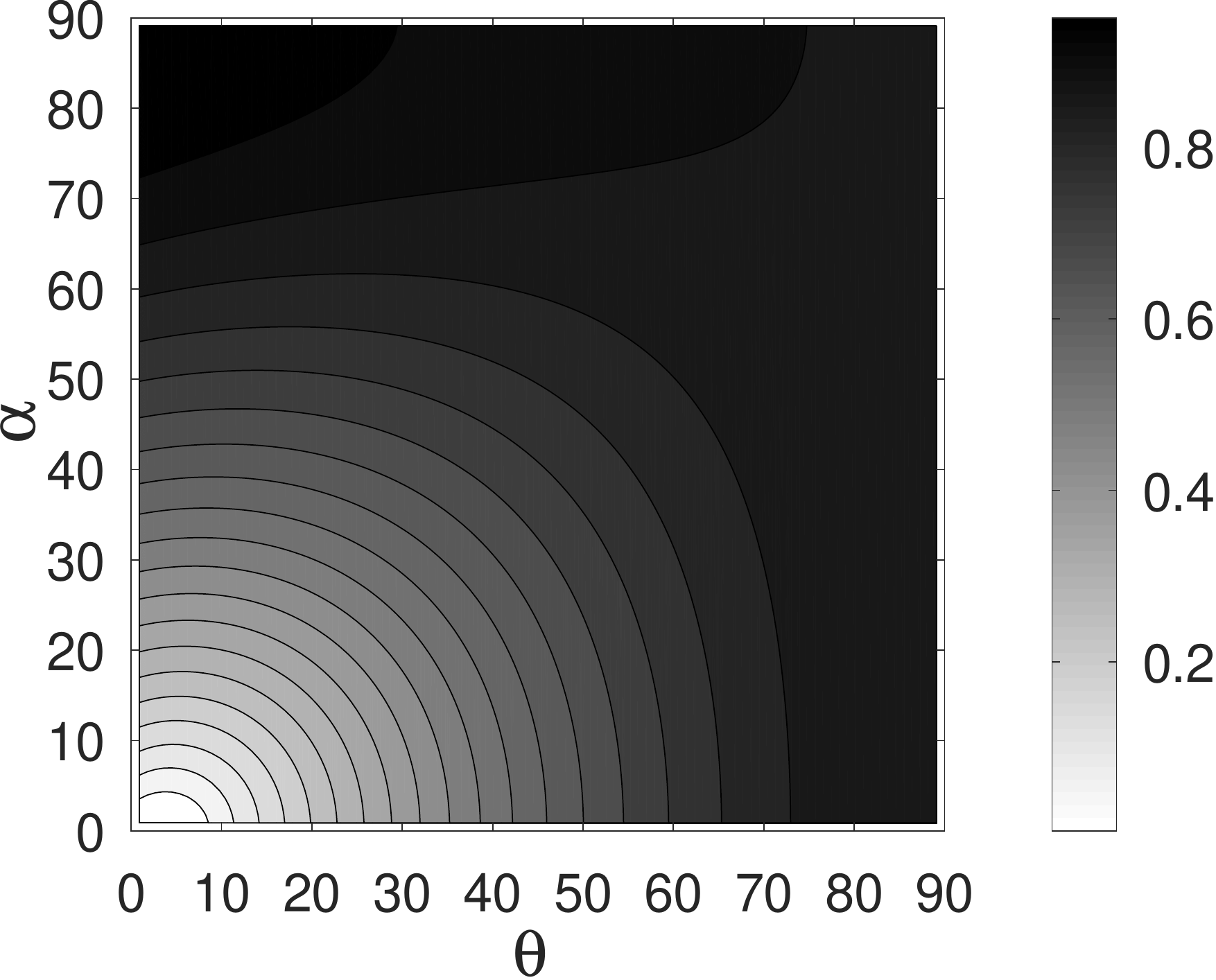}}
\subfigure[]{\includegraphics[scale=0.28]{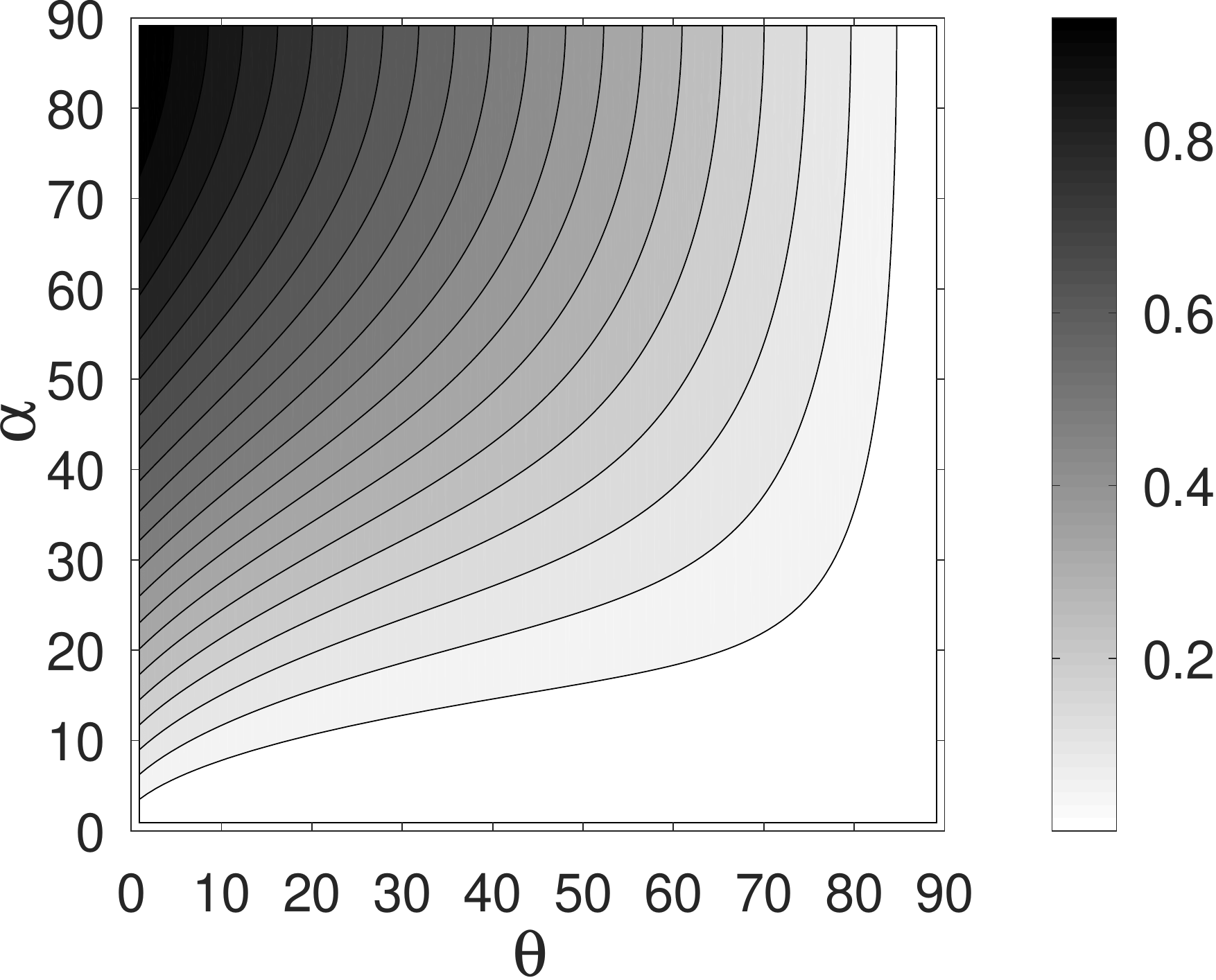}}
\subfigure[]{\includegraphics[scale=0.28]{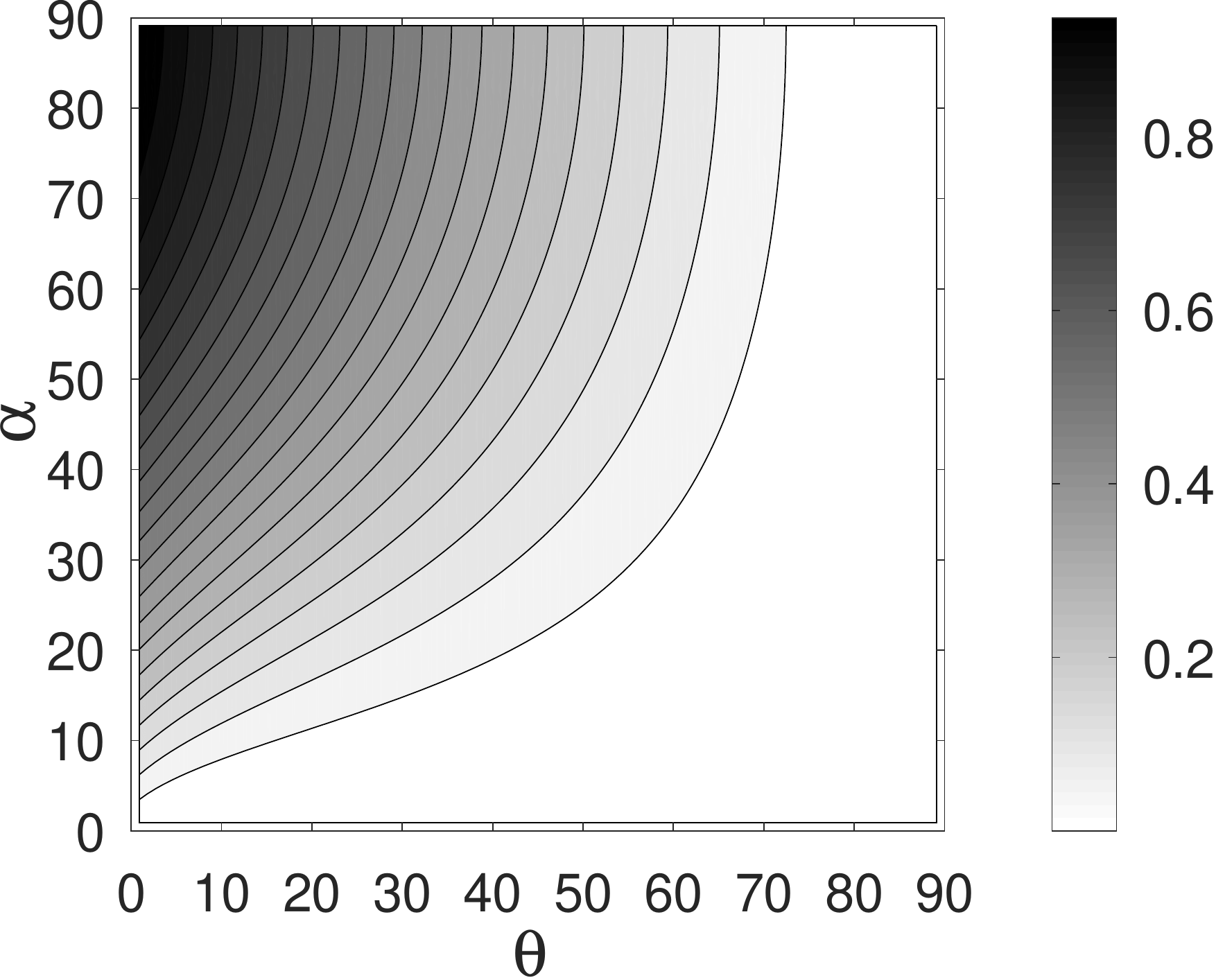}}
\caption{Contours of ($|\omega_0|-|\omega_1|)/2\Omega$. From left to right $N/2\Omega=0.1$, 1 and 10.}\label{fig2}
\end{figure}

In Case 2, we can further discuss the energy ratios of reflected and transmitted waves. Take Configuration 1 for an example. The wave solutions are
\begin{equation}
\left\{
\begin{aligned}
\mbox{incident wave: }&u_z'=\Re\left\{a_2\exp\left[i(k_x x+k_y y+q_2z-\omega t)\right]\right\} \\
\mbox{reflected wave: }&u_z'=\Re\left\{a_1\exp\left[i(k_x x+k_y y+q_1z-\omega t)\right]\right\} \\
\mbox{transmitted wave: }&u_z'=\Re\left\{b_2\exp\left[i(k_x x+k_y y+s_2z-\omega t)\right]\right\}
\end{aligned}
\right.
\end{equation}
where $q_{1,2}$ and $s_{1,2}$ depend on $\theta$ and $\alpha$,
\begin{align}
q_{1,2}&=\frac{k}{4\Omega^2\sin^2\theta-\omega^2}\left(-4\Omega^2\sin\theta\cos\theta\sin\alpha\pm\sqrt{\omega^2(\omega_0^2-\omega^2)}\right), \\
s_{1,2}&=\frac{k}{4\Omega^2\sin^2\theta-\omega^2}\left(-4\Omega^2\sin\theta\cos\theta\sin\alpha\pm\sqrt{\omega^2(\omega_0^2-\omega^2+N^2)-4\Omega^2N^2\sin^2\theta}\right).
\end{align}
At $\theta=90^\circ$ the above two expressions of $q$ and $s$ reduce to \eqref{q} and \eqref{s}. By the boundary conditions \eqref{b1} and \eqref{b2} we can derive
\begin{equation}
a_1=\frac{s_2-q_2}{q_1-s_2}a_2, \; b_2=\frac{q_1-q_2}{q_1-s_2}a_2.
\end{equation}
Similar to \eqref{uxuy1} we can derive
\begin{equation}\label{uxuy2}
\begin{aligned}
\hat u_x&=\frac{2\Omega k_y^2\cos\theta}{i\omega k^2}\hat u_z-\frac{\omega k_x+2i\Omega k_y\sin\theta}{i\omega k^2}D\hat u_z, \\
\hat u_y&=-\frac{2\Omega k_xk_y\cos\theta}{i\omega k^2}\hat u_z+\frac{2i\Omega k_x\sin\theta-\omega k_y}{i\omega k^2}D\hat u_z.
\end{aligned}
\end{equation}
Substitution of \eqref{uxuy2} into \eqref{energy} leads to the reflection ratio
\begin{equation}\label{ratio5}
\begin{aligned}
&\frac{a_1^2}{a_2^2}\cdot\frac{4\Omega^2(k_y\cos\theta+q_1\sin\theta)^2+\omega^2(k^2+q_1^2)}{4\Omega^2(k_y\cos\theta+q_2\sin\theta)^2+\omega^2(k^2+q_2^2)} \\
=&\left(\frac{s_2-q_2}{q_1-s_2}\right)^2\cdot\frac{4\Omega^2(k_y\cos\theta+q_1\sin\theta)^2+\omega^2(k^2+q_1^2)}{4\Omega^2(k_y\cos\theta+q_2\sin\theta)^2+\omega^2(k^2+q_2^2)}.
\end{aligned}
\end{equation}
At $\theta=90^\circ$, $q_1=q$ and $q_2=-q$ the reflection ratio reduces to \eqref{ratio1}. We now investigate \eqref{ratio5}. Firstly, the reflection ratio is independent of the magnitude $k$ of wave vector, because $q_{1,2}$ and $s_{1,2}$ are all proportional to $k$, but depends on the direction $\alpha$ of wave vector. Secondly, we have already known that to support wave propagation across interface it is required that $\omega_1^2<\omega^2<\omega_0^2$, and so we take the average of reflection ratio with $\omega^2$ uniformly spacing between $\omega_1^2$ and $\omega_0^2$. Consequently, the reflection ratio depends on $\theta$, $\alpha$ and $N/2\Omega$. 

\begin{figure}
\centering
\subfigure[]{\includegraphics[scale=0.28]{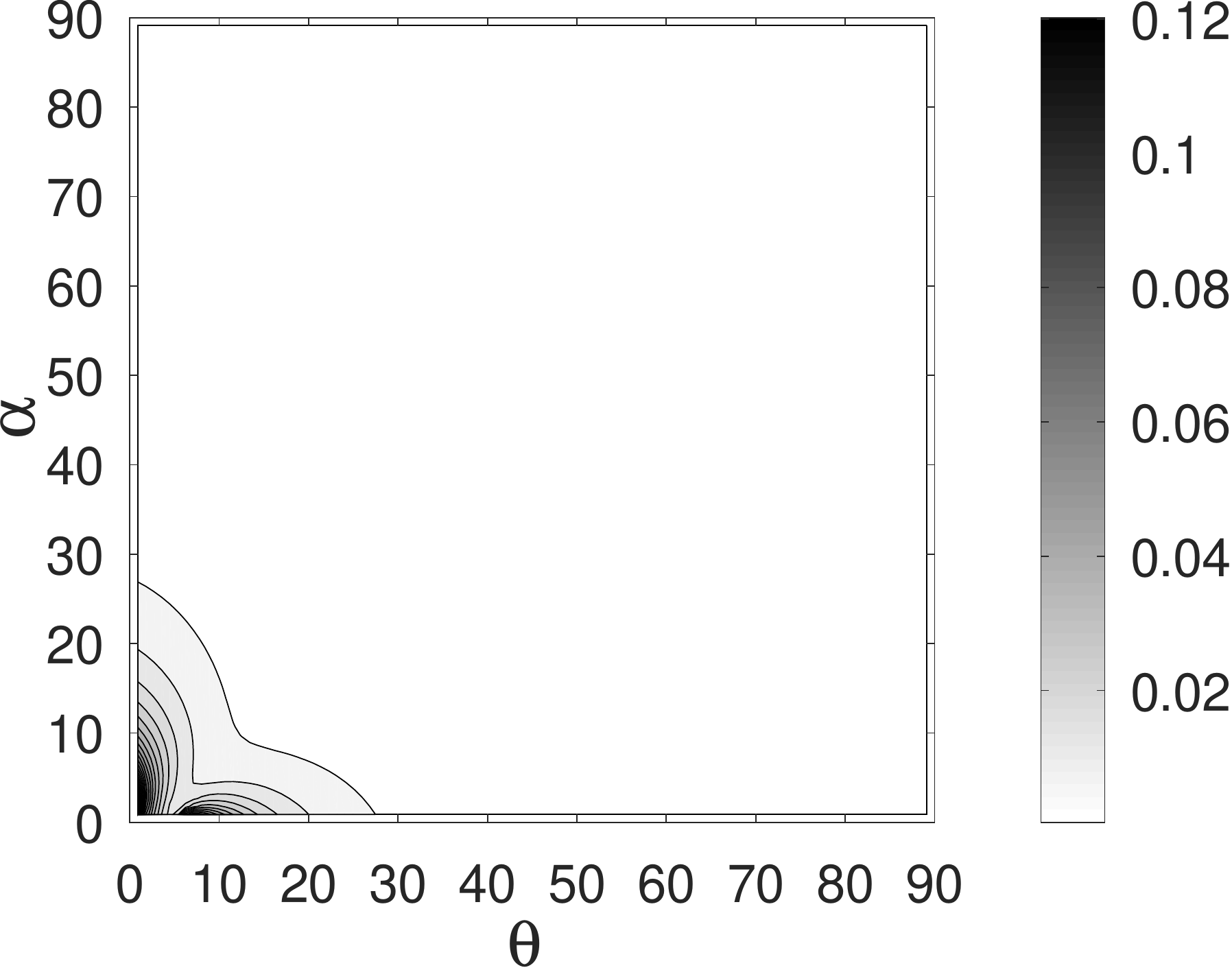}}
\subfigure[]{\includegraphics[scale=0.28]{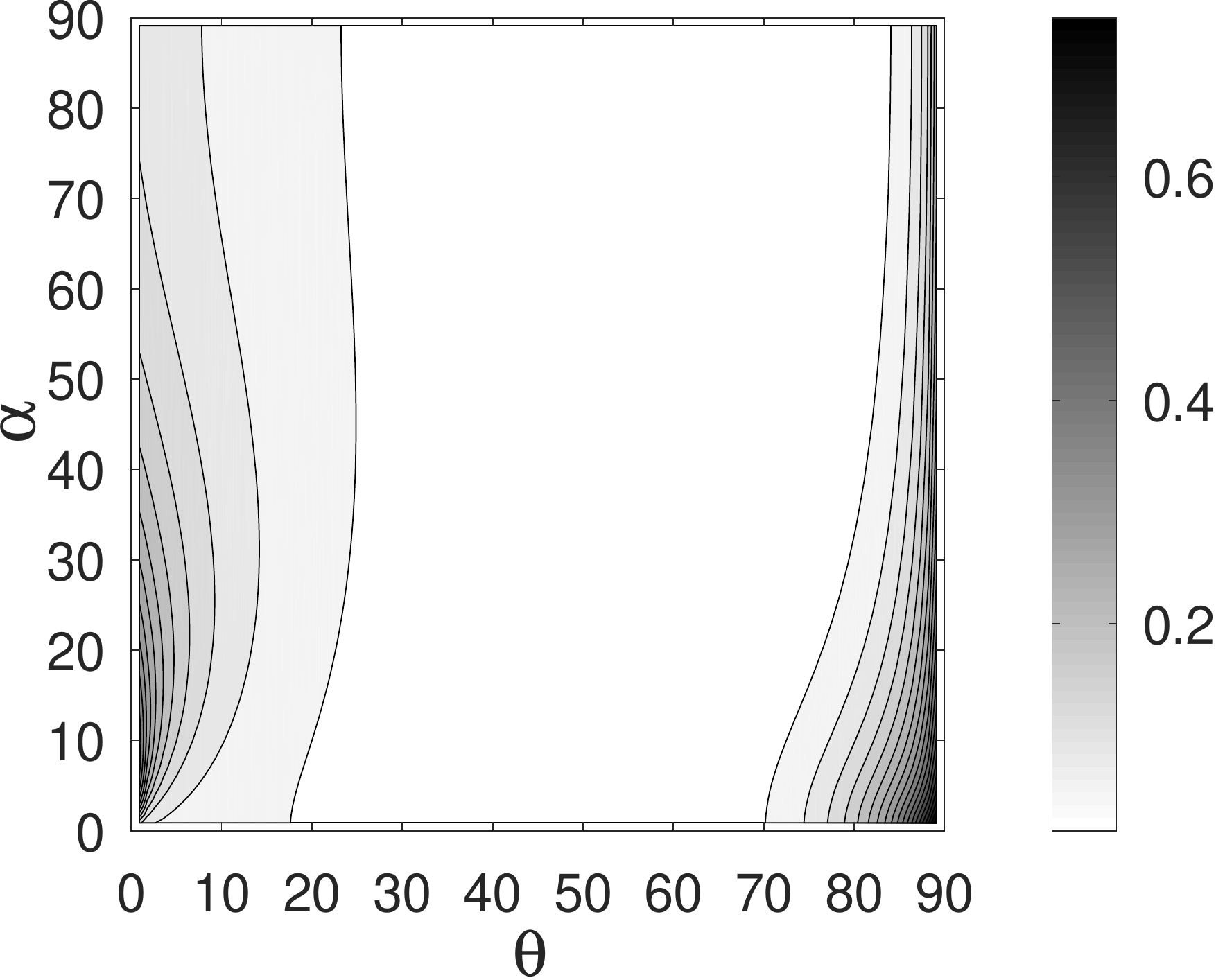}}
\subfigure[]{\includegraphics[scale=0.28]{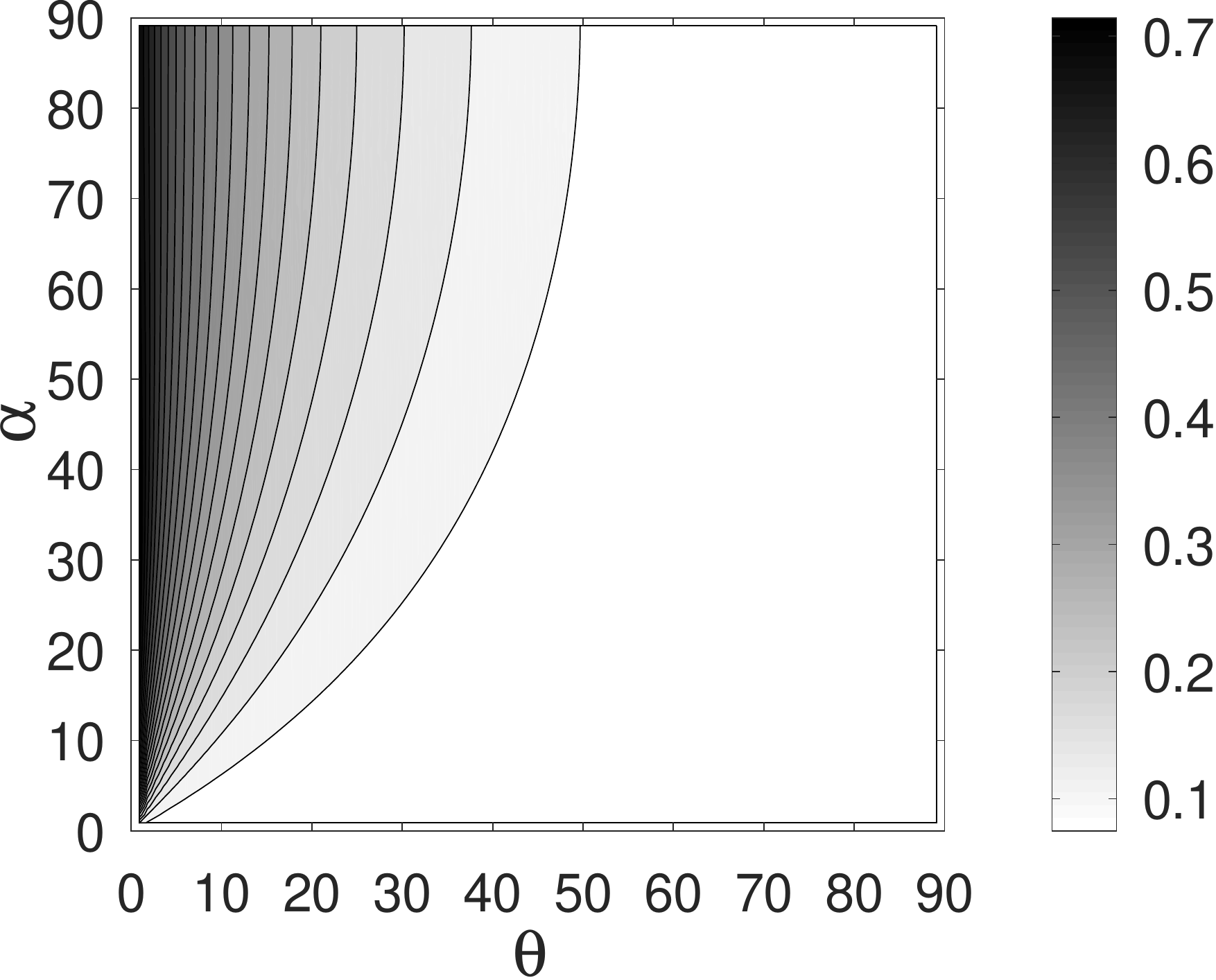}}
\caption{Contours of reflection ratio of Configuration 1. From left to right $N/2\Omega=0.1$, 1 and 10.}\label{fig3}
\end{figure}

Figure \ref{fig3} shows the contours of reflection ratio versus $\theta$ and $\alpha$ at $N/2\Omega=0.1$, 1 and 10. It provides two results. The first is about fast rotation at $N/2\Omega=0.1$. The small reflection ratio in panel (a) indicates that fast rotation indeed favours wave transmission as already found at $\theta=90^\circ$ with $N/2\Omega<1$. The second is about slow rotation at $N/2\Omega=10$. Panel (c) of Figure \ref{fig2} implies that at slow rotation the equatorial area may possibly support wave propagation because this area has a wide waveband for wave propagation. However, panel (c) of Figure \ref{fig3} denies this possibility. The dark area in the low-latitude area indicates that most of wave is reflected but not transmitted. Instead, it indicates that wave transmission may occur in the high-latitude area. Nevertheless, in the high-latitude area only a narrow band of wave is allowed to transmit so that the total transmission cannot be strong. In a word, slow rotation inhibits wave transmission.

\begin{figure}
\centering
\subfigure[]{\includegraphics[scale=0.28]{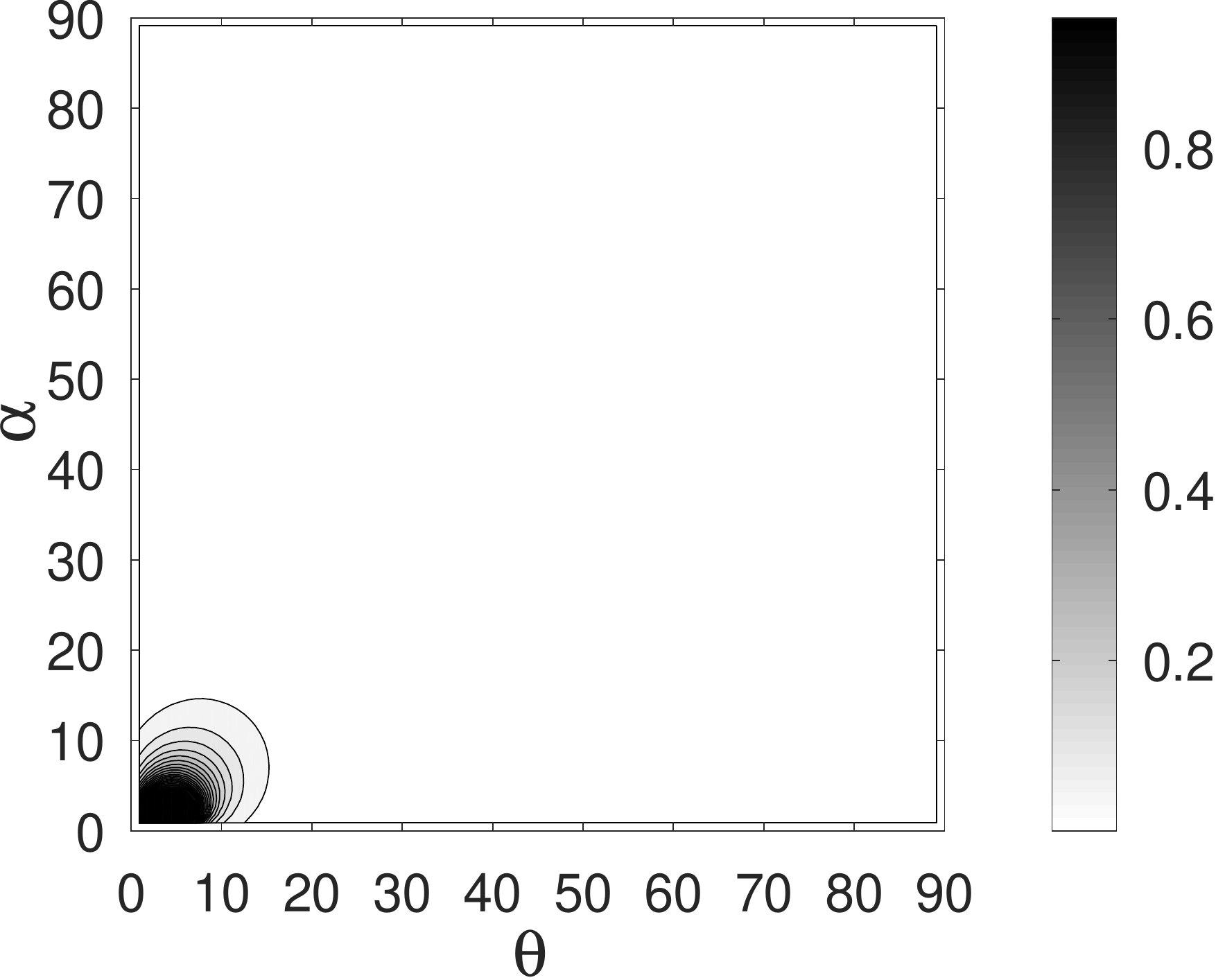}}
\subfigure[]{\includegraphics[scale=0.28]{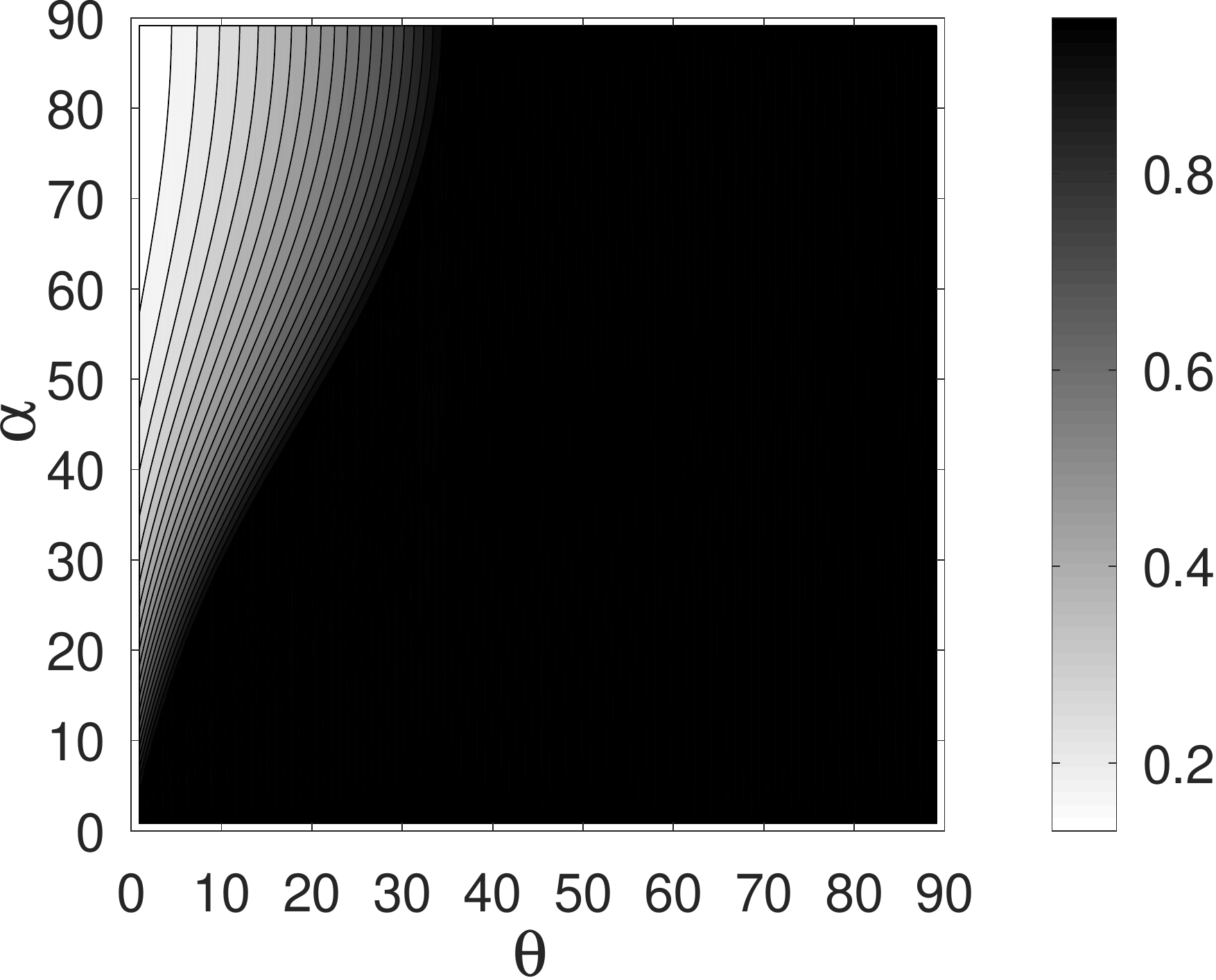}}
\subfigure[]{\includegraphics[scale=0.28]{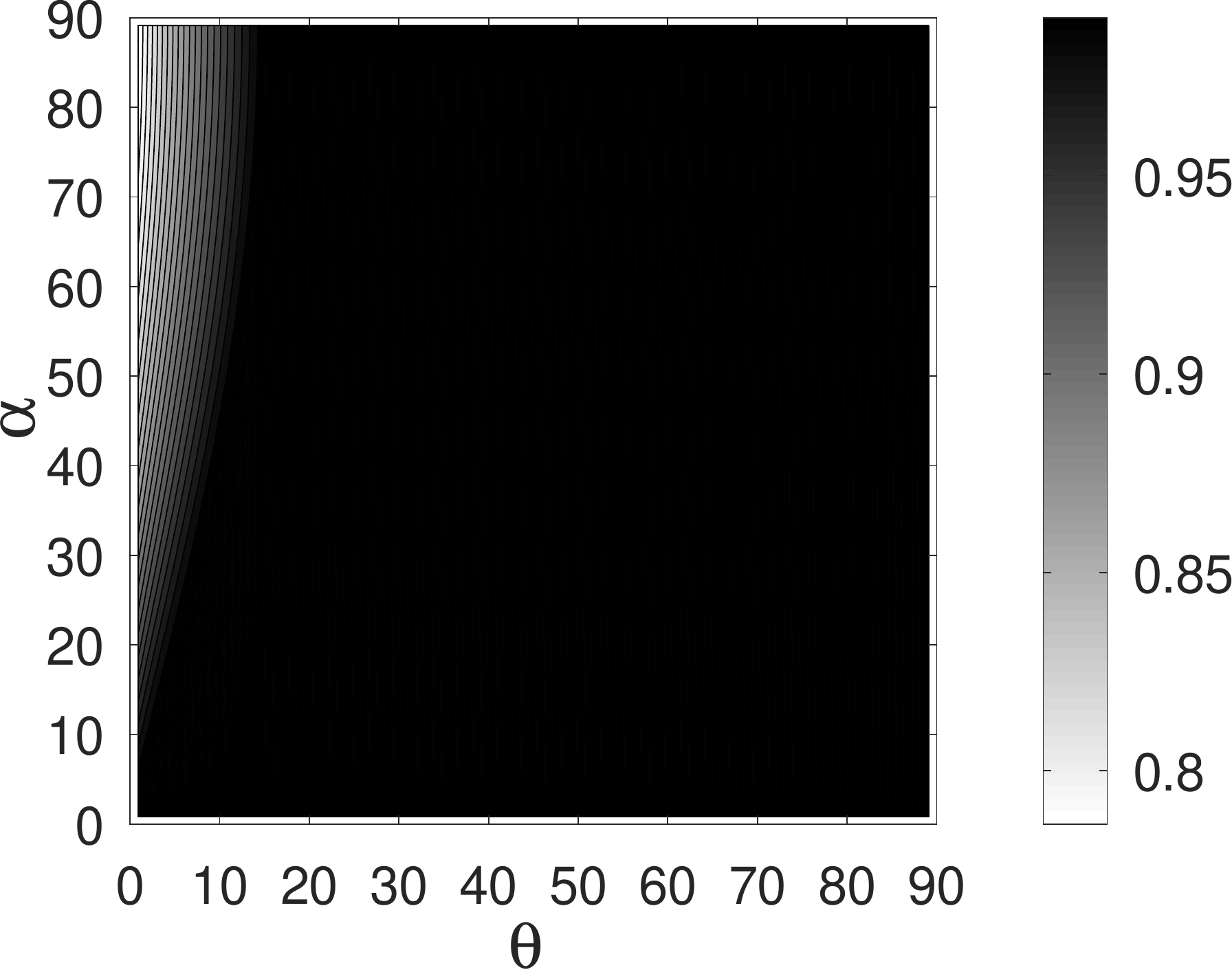}}
\caption{Contours of reflection ratio of Configuration 2. From left to right $N/2\Omega=0.1$, 1 and 10.}\label{fig4}
\end{figure}

\begin{figure}
\centering
\subfigure[]{\includegraphics[scale=0.28]{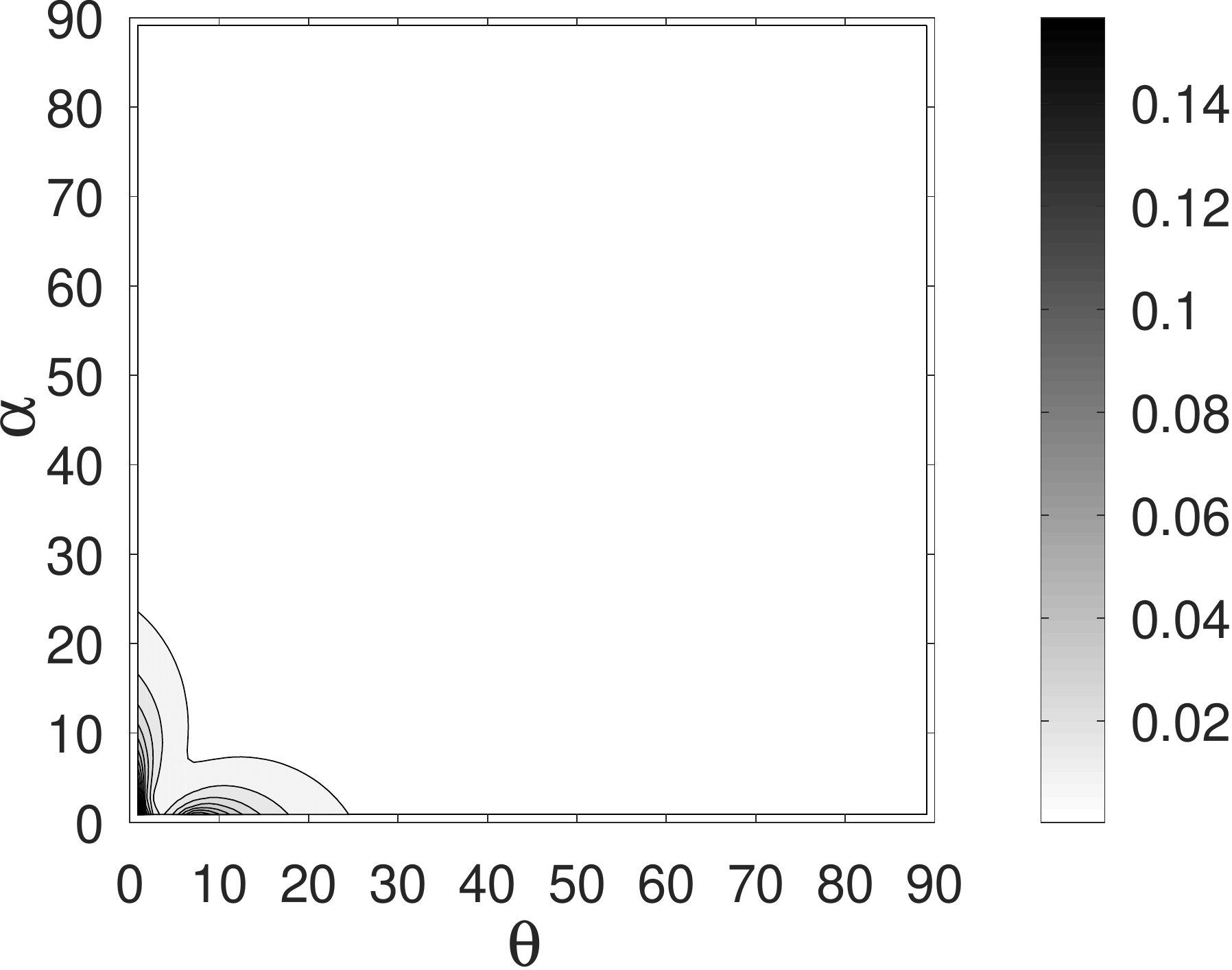}}
\subfigure[]{\includegraphics[scale=0.28]{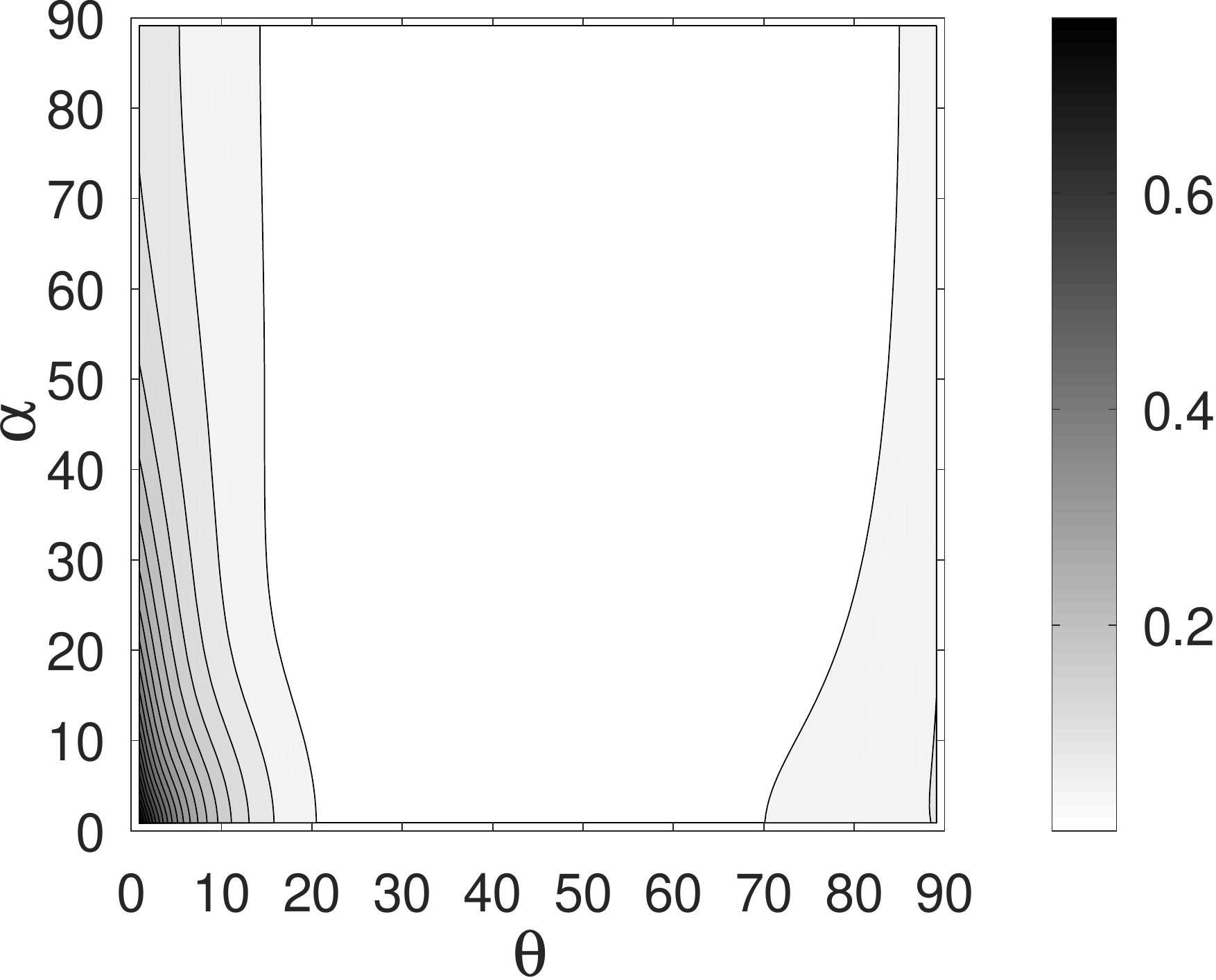}}
\subfigure[]{\includegraphics[scale=0.28]{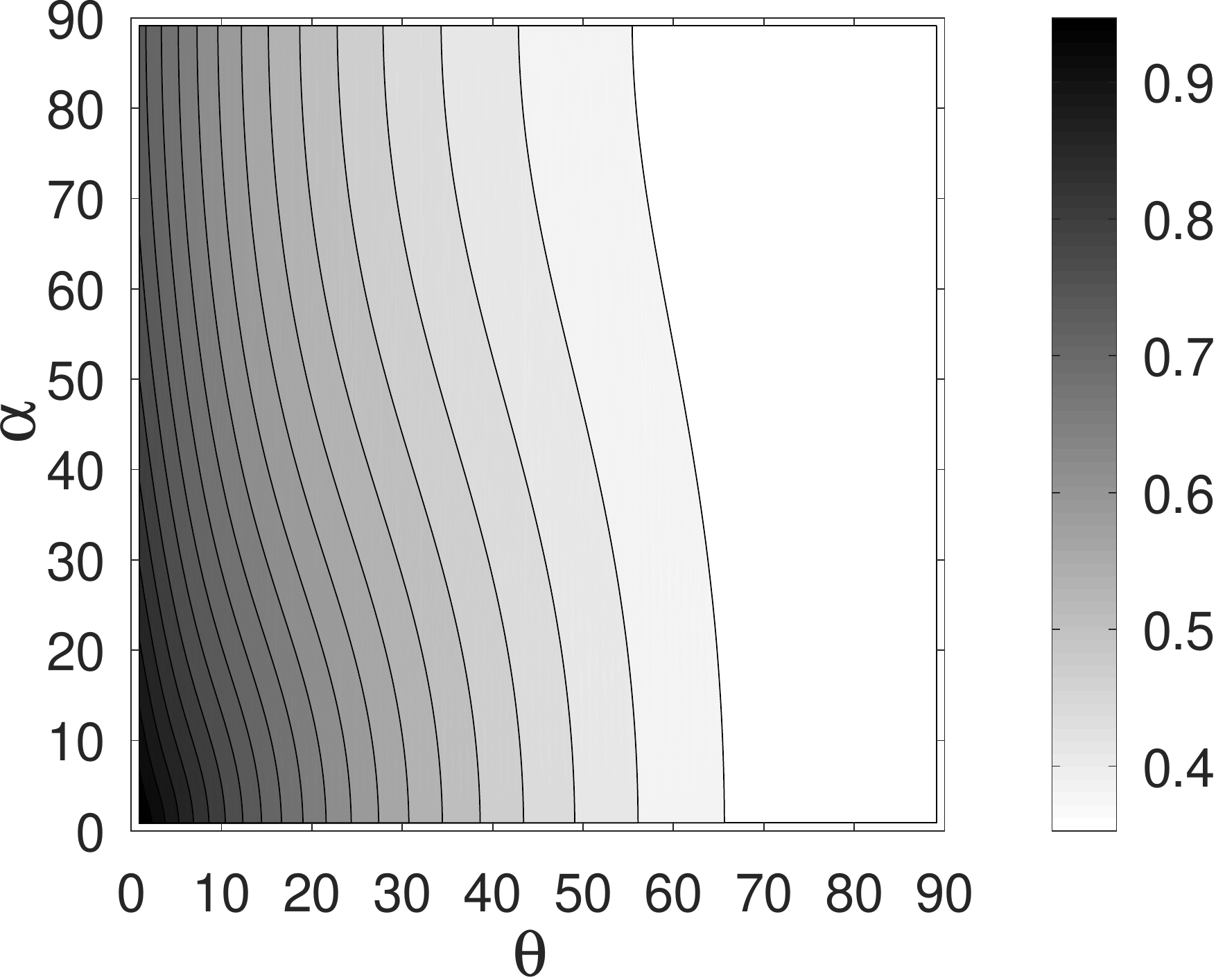}}
\caption{Contours of reflection ratio of Configuration 3. From left to right $N/2\Omega=0.1$, 1 and 10.}\label{fig5}
\end{figure}

\begin{figure}
\centering
\subfigure[]{\includegraphics[scale=0.28]{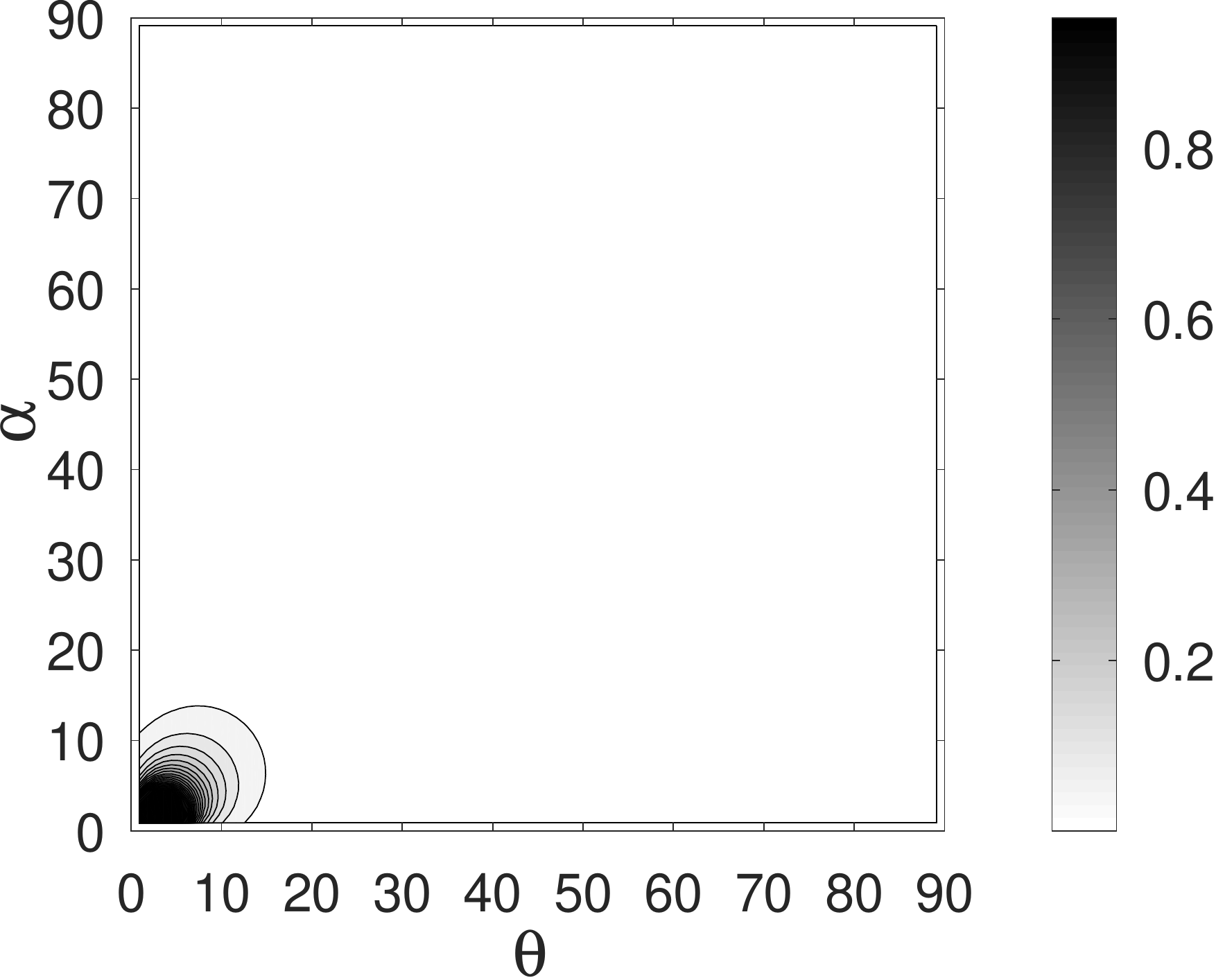}}
\subfigure[]{\includegraphics[scale=0.28]{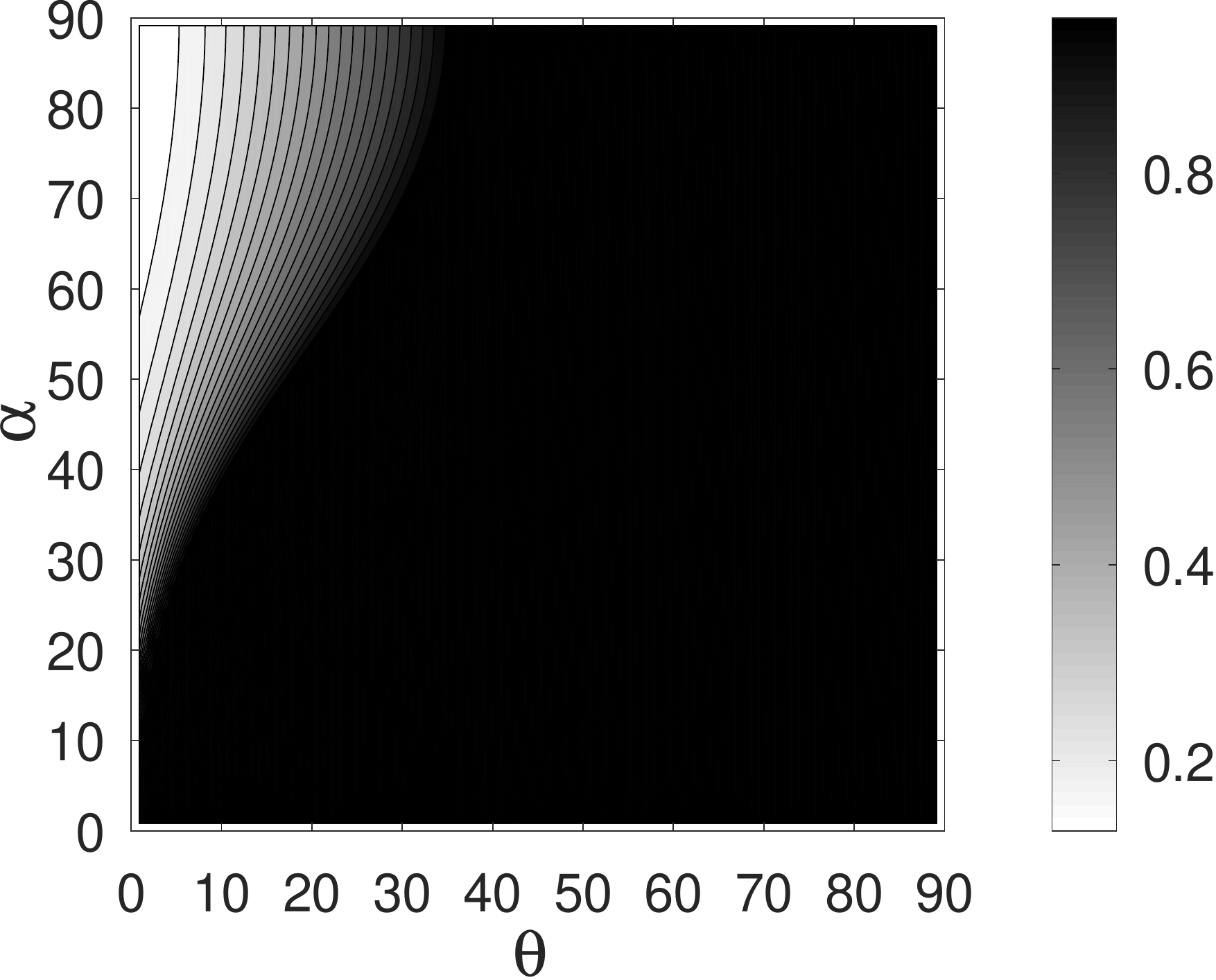}}
\subfigure[]{\includegraphics[scale=0.28]{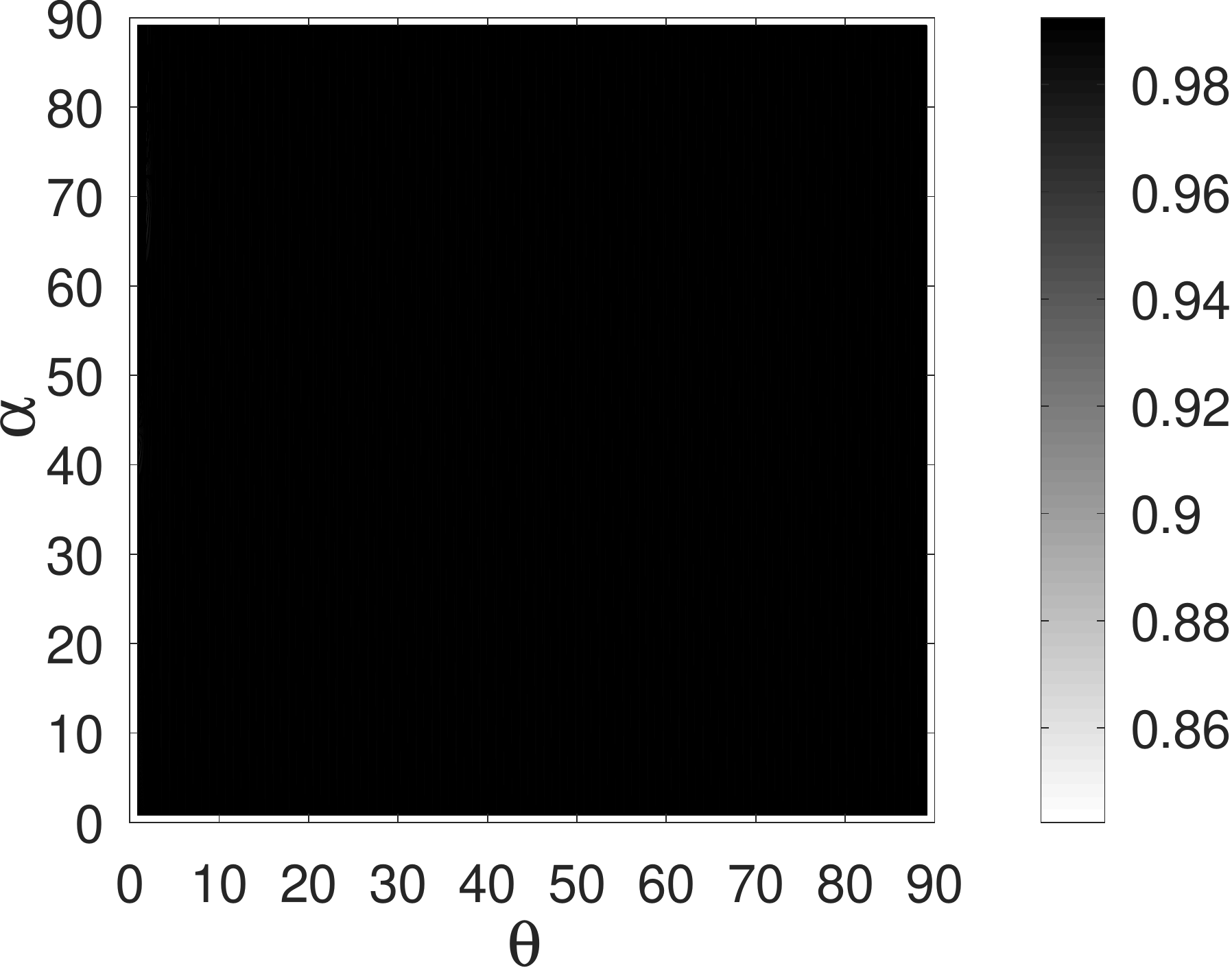}}
\caption{Contours of reflection ratio of Configuration 4. From left to right $N/2\Omega=0.1$, 1 and 10.}\label{fig6}
\end{figure}

Similarly, the reflection ratio of Configuration 2 is
\begin{equation}
\left(\frac{s_1-q_1}{q_1-s_2}\right)^2\cdot\frac{4\Omega^2(k_y\cos\theta+s_2\sin\theta)^2+\omega^2(k^2+s_2^2)}{4\Omega^2(k_y\cos\theta+s_1\sin\theta)^2+\omega^2(k^2+s_1^2)},
\end{equation}
that of Configuration 3 is
\begin{equation}
\left(\frac{q_2-s_2}{s_1-q_2}\right)^2\cdot\frac{4\Omega^2(k_y\cos\theta+s_1\sin\theta)^2+\omega^2(k^2+s_1^2)}{4\Omega^2(k_y\cos\theta+s_2\sin\theta)^2+\omega^2(k^2+s_2^2)},
\end{equation}
and that of Configuration 4 is
\begin{equation}
\left(\frac{q_1-s_1}{s_1-q_2}\right)^2\cdot\frac{4\Omega^2(k_y\cos\theta+q_2\sin\theta)^2+\omega^2(k^2+q_2^2)}{4\Omega^2(k_y\cos\theta+q_1\sin\theta)^2+\omega^2(k^2+q_1^2)}.
\end{equation}
At some $\theta$ and $\alpha$, especially with strong stratification $N/2\Omega=10$, the reflection ratio is calculated to be greater than 1. This might arise from the Boussinesq approximation in which constant $\rho$ in perturbation equations brings an error when stratification is strong. We then set the threshold at 1. The contours of the three reflection ratios are shown in Figures \ref{fig4}, \ref{fig5} and \ref{fig6}.  These figures provide two results. The first is that the symmetry at $\theta=90^\circ$ is lost, namely the reflection ratios of Configurations 1 and 4 or of Configurations 2 and 3 are no longer identical. The second is that, again, fast rotation at $N/2\Omega=0.1$ favours wave transmission whereas slow rotation at $N/2\Omega=10$ inhibits wave transmission. Slow rotation of Configuration 2 (Figure \ref{fig4}) corresponds to the detection of g mode on the solar surface. Therefore, in the slowly rotating Sun ($N/2\Omega$ at the order of 10 to 100) the gravito-inertial wave in the radiative region seems unlikely to transmit to the convective region.

\section{Discussions}
Our results may provide guidance for the detection of modes, e.g. modes in the inner region of a slowly rotating star or planet are hardly detected on the surface. Our results may be also useful for stellar oscillation observations. For a rapidly rotating solar-type star, gravito-inertial wave can propagate outward to convective region such that we can know more about the structure of the inner radiative region. Similarly, for a rapidly rotating massive star, inertial wave can propagate outward to radiative region such that we can know more about the differential rotation of the inner convective region. Moreover, since wave can propagate across interface to the other region, angular momentum and energy can be carried out farther away than expected before. For example, in \citep{goldreich}, in a non-rotating early-type star, it was believed that gravity wave excited at interface propagates towards the stellar surface and damps just below photosphere. According to our model, much of wave can propagate in an opposite way towards the convective region provided that star rotates sufficiently fast.

Certainly, our simplified model should be numerically validated. One can perform numerical calculations in a global spherical geometry by imposing interface with an overshooting profile \citep{zhang,mathis}. Anelastic \citep{rogers} or fully compressible fluid can be used for numerical study, the latter will bring numerical difficulty due to fast sound speed. In addition, magnetic field plays an important role in wave propagation, which radically changes the wave dispersion relation \citep{wei16,wei18,lin}, so that the magnetic effect on wave reflection and transmission also needs to be studied.

\section*{Acknowledgements}
An anonymous referee carefully examined my calculations and gave me many good suggestions. I am financially supported by 1000 youth talents program of Chinese government and National Natural Science Foundation of China (grant no. 11872246).

\bibliographystyle{apj}
\bibliography{paper}

\newpage
\begin{center}\Large\bf Erratum\end{center}

An error is found in calculating the wave reflection and transmission ratios at the interface. I used kinetic energy to calculate the ratios such that in some regions reflection ratio appears to be greater than 1, which is unphysical. To find the correct physical quantity we derive the total energy equation. We perform the dot product $\bm u'\cdot$ with the momentum perturbation equation (note that $\bm u'\cdot$ Coriolis force vanishes), multiply $(g^2/N^2\rho^2)\rho'$ with the density perturbation equation (only in the stratified region), and add them together. Thus, the total energy equation is derived to be
\begin{equation}
\frac{\partial}{\partial t}\left(\frac{u'^2}{2}+\frac{g^2}{N^2\rho^2}\frac{\rho'^2}{2}\right)=-\frac{1}{\rho}\bm\nabla\cdot(p'\bm u'),
\end{equation}
where in bracket of LHS the first term is kinetic energy per unit mass and the second term is the buoyancy energy per unit mass, and the term in bracket of RHS is energy flux. It should be noted that this equation is for the stratified region. In the convective region it is simply that the buoyancy energy is zero. Through the total energy equation, we can find that the correct quantity to measure the reflection and transmission ratios should be energy flux $\langle p'\bm u'\rangle$. Moreover, the boundary condition shows that the vertical energy flux $\langle p'u_z'\rangle$ of incident wave is equal to the sum of those of reflected and transmitted waves. Therefore, with the vertical energy flux to measure ratios, the sum of two ratios is equal to 1. Next, we calculate the vertical energy flux $\langle p'u_z'\rangle$. We express $p'=\Re\{\tilde p'\}$ and $u_z'=\Re\{\tilde u_z'\}$ where tilde denotes the complex variables of perturbations. The complex pressure perturbation $\tilde p'$ can be calculated from perturbation equations (40) or (41) to be
\begin{align}
\begin{aligned}
i\omega k^2\frac{\tilde p'}{\rho}&=2\Omega\cos\theta(2i\Omega k_y\sin\theta-\omega k_x)\tilde u_z'+(4\Omega^2\sin^2\theta-\omega^2)\frac{\partial \tilde u_z'}{\partial z} \\
&=\left[-2\Omega\omega k\cos\theta\cos\alpha+i(4\Omega^2k\cos\theta\sin\theta\sin\alpha+4\Omega^2\gamma\sin^2\theta-\gamma\omega^2)\right]\tilde u_z',
\end{aligned}
\end{align}
where $\gamma$ is the vertical wavenumber, i.e. $q_{1,2}$ or $s_{1,2}$. The time-averaged vertical energy flux $\langle p'u_z'\rangle$ is then
\begin{equation}
\langle p'u_z'\rangle=\langle \Re\{\tilde p'\}\Re\{\tilde u_z'\}\rangle=\frac{1}{2}\rho|\tilde u_z'|^2\frac{4\Omega^2}{\omega k}\left[\cos\theta\sin\theta\sin\alpha+\frac{\gamma}{k}(\sin^2\theta-\frac{\omega^2}{4\Omega^2})\right].
\end{equation} 
Interestingly, by the expressions of $\gamma$ (Eqs. (51) and (52)), the reflection ratio $\eta=|\langle p'u_z'\rangle_r/\langle p'u_z'\rangle_i|$ in the four configurations is identical, 
\begin{equation}
%eta=\left(\frac{q_2-s_2}{q_1-s_2}\right)^2, \hspace{2mm} \left(\frac{q_1-s_1}{q_1-s_2}\right)^2, \hspace{2mm} \left(\frac{q_2-s_2}{q_2-s_1}\right)^2, \hspace{2mm} \left(\frac{q_1-s_1}{q_2-s_1}\right)^2 
\eta=\left(\frac{\sqrt{\omega^2(\omega_0^2-\omega^2)}-\sqrt{\omega^2(\omega_0^2-\omega^2+N^2)-4\Omega^2N^2\sin^2\theta}}{\sqrt{\omega^2(\omega_0^2-\omega^2)}+\sqrt{\omega^2(\omega_0^2-\omega^2+N^2)-4\Omega^2N^2\sin^2\theta}}\right)^2.
\end{equation}
The transmission ratio is $|\langle p'u_z'\rangle_t/\langle p'u_z'\rangle_i|=1-\eta$. It is reasonable that the partition of energy flux at the interface depends on the wave properties, i.e. frequency and wavevector, and medium properties, i.e. rotation rate and stratification strength, so that the four configurations have the same ratios. The above expression shows that the two ratios depend on 4 parameters: $N/2\Omega$, $\omega/2\Omega$, latitude $\theta$, and the direction of the wavevector $\alpha$ ($\omega_0$ depends on $\alpha$, see Eq. (47)). We take three values of $N/2\Omega=(0.1, 1, 10)$ and three values of $\sin^2\alpha=(0.1, 0.5, 0.9)$ to plot the contours of the reflection ratio $\eta$ versus wave frequency $\omega/2\Omega$ and latitude $\theta$ in Figure 1. In each panel, wave exists between the top and bottom curves, i.e. $\omega_1<\omega<\omega_0$, and outside the two curves wave cannot propagate across the interface. At each row, panels from left to right show that stronger stratification or slower rotation leads to stronger reflection and hence weaker transmission. At $N/2\Omega=10$ wave hardly transmits, at $N/2\Omega=1$ wave transmits at lower latitudes with a wider waveband, and at $N/2\Omega=0.1$ wave transmits at any latitudes with a wide waveband. This result is consistent with the paper, i.e. rotation favours transmission whereas stratification inhibits transmission. In each column, panels from top to bottom show that larger $\alpha$, i.e. wavevectors that are aligned with the north-south direction rather than the east-west direction, lead to a wider waveband of transmission. 
\begin{figure}
\centering
\subfigure[]{\includegraphics[scale=0.3]{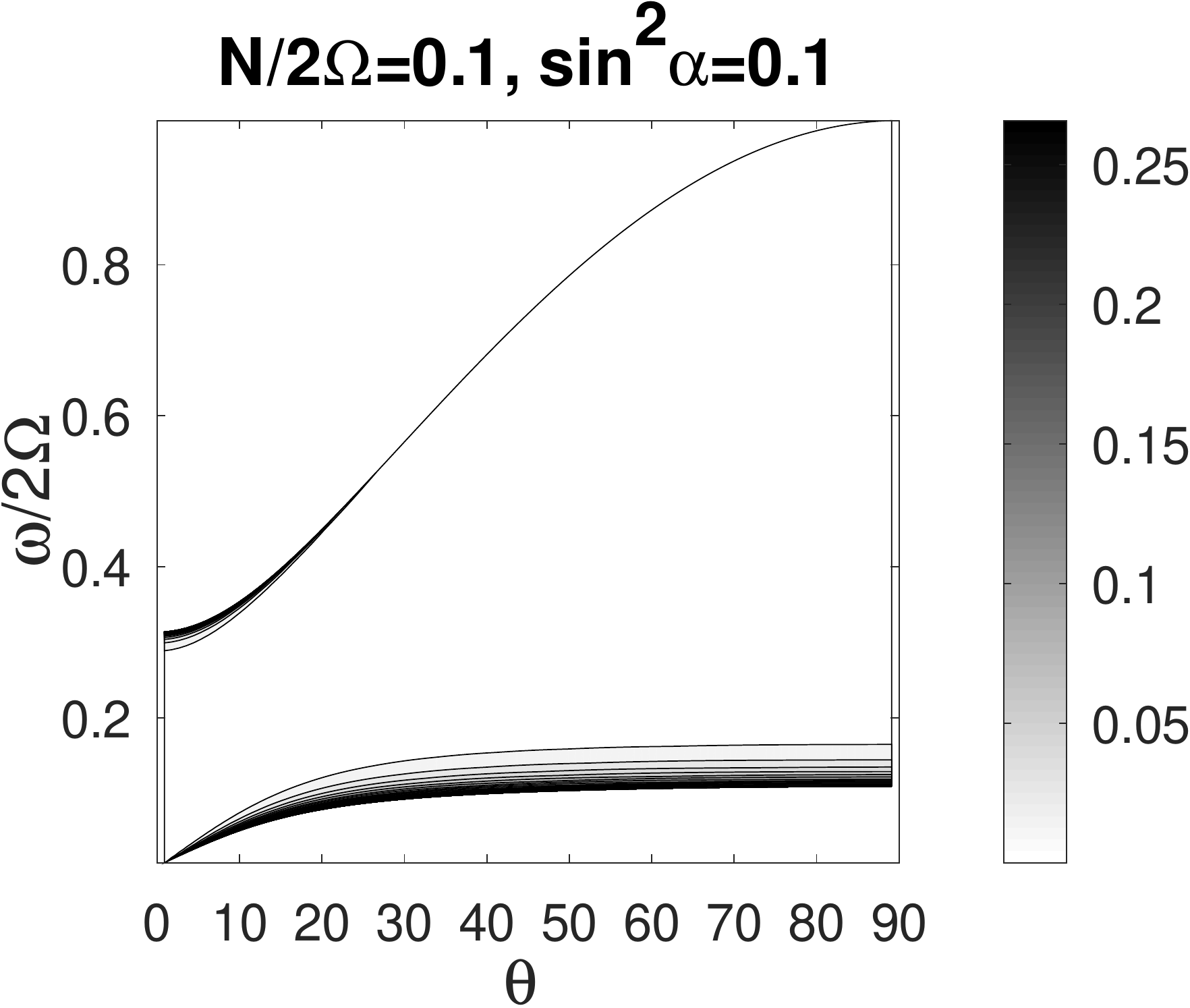}}
\subfigure[]{\includegraphics[scale=0.3]{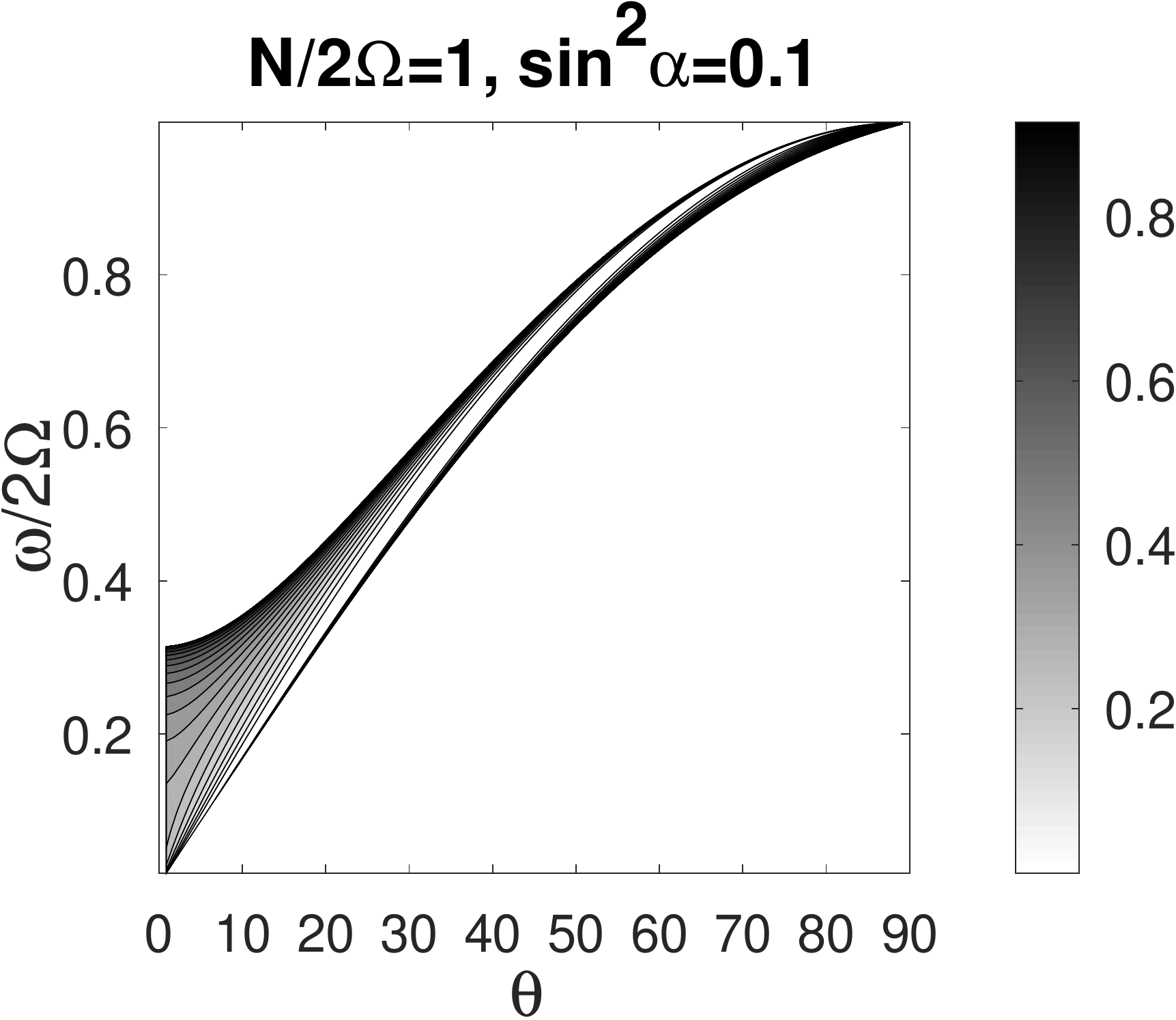}}
\subfigure[]{\includegraphics[scale=0.3]{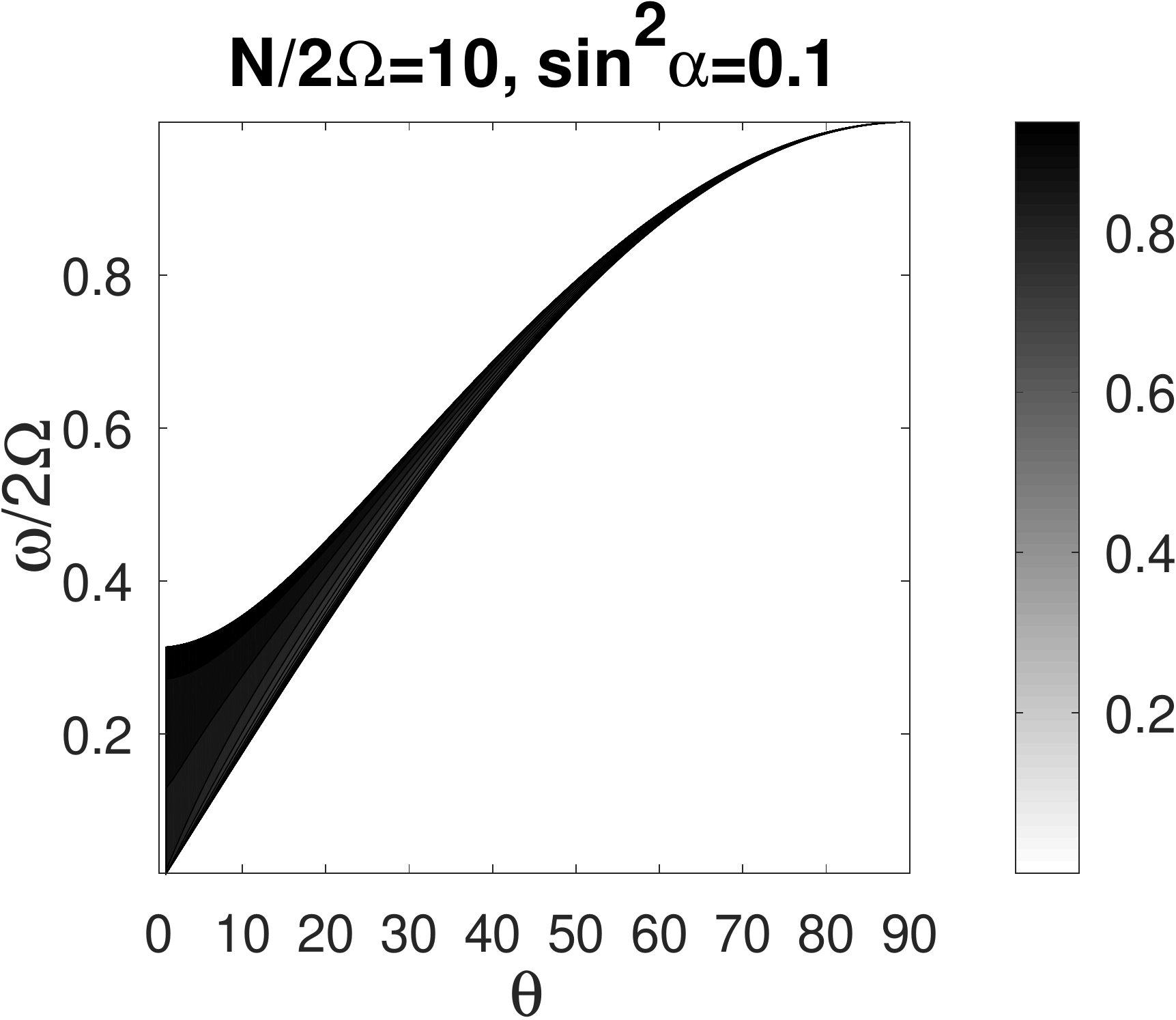}}
\subfigure[]{\includegraphics[scale=0.3]{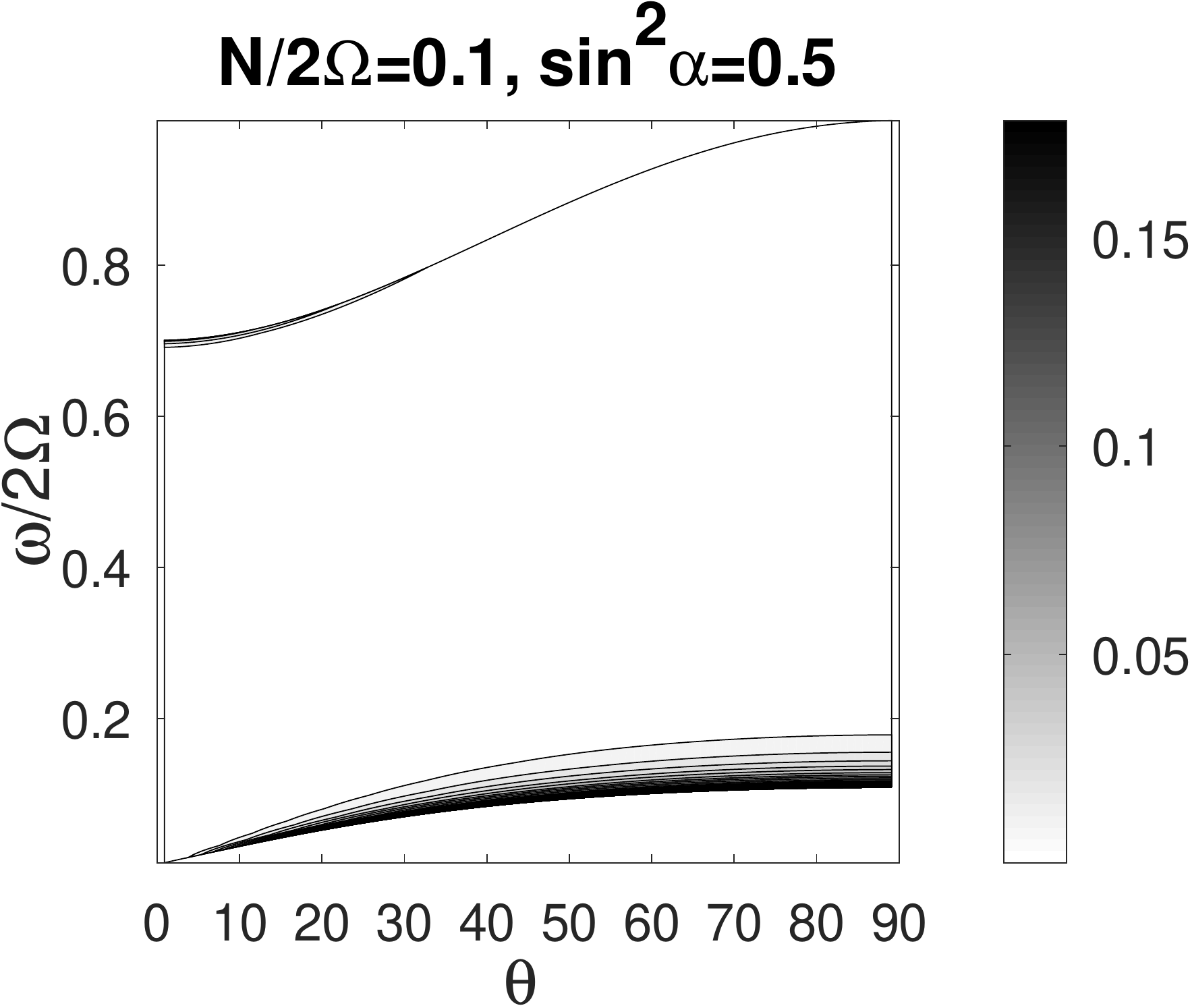}}
\subfigure[]{\includegraphics[scale=0.3]{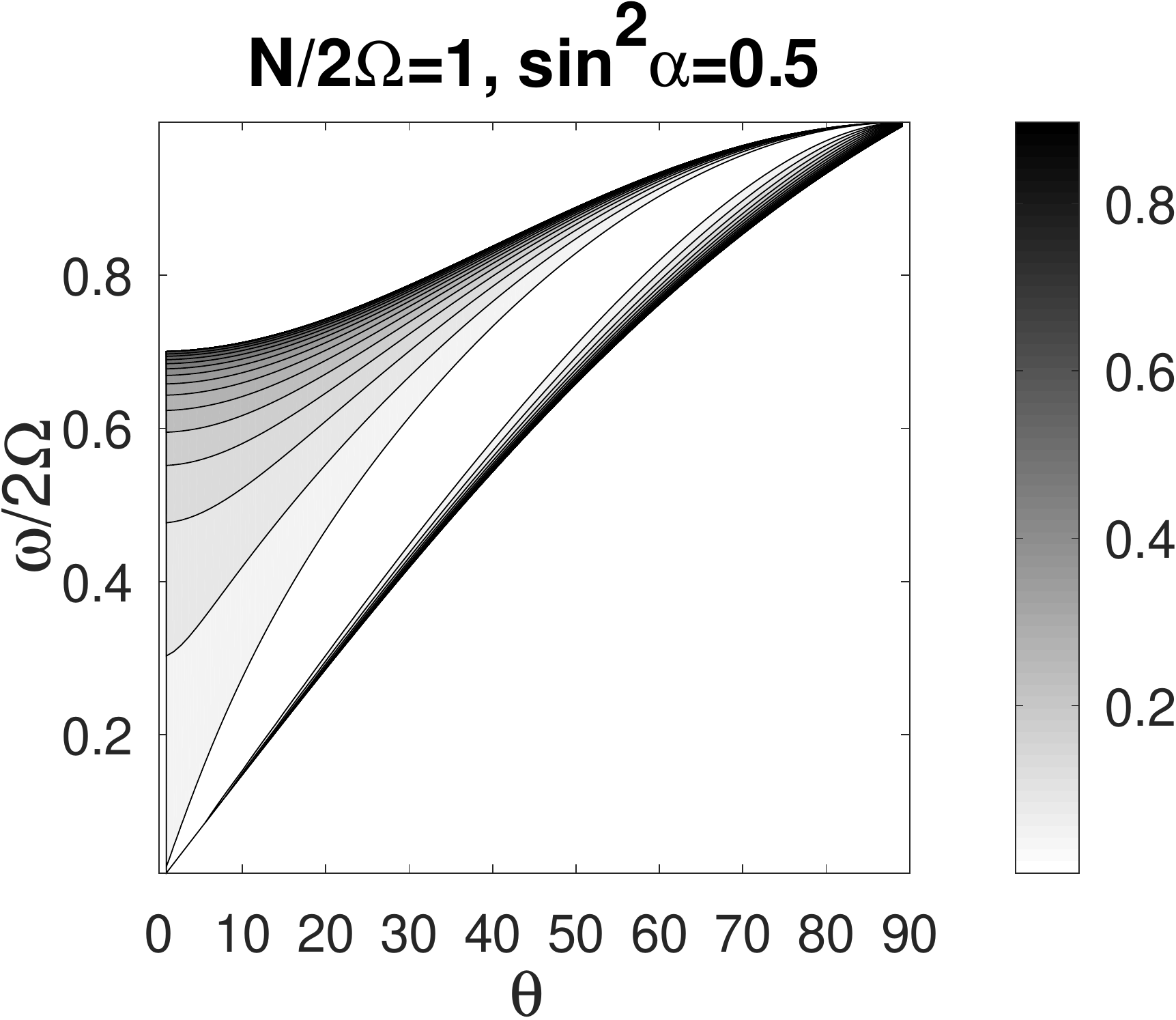}}
\subfigure[]{\includegraphics[scale=0.3]{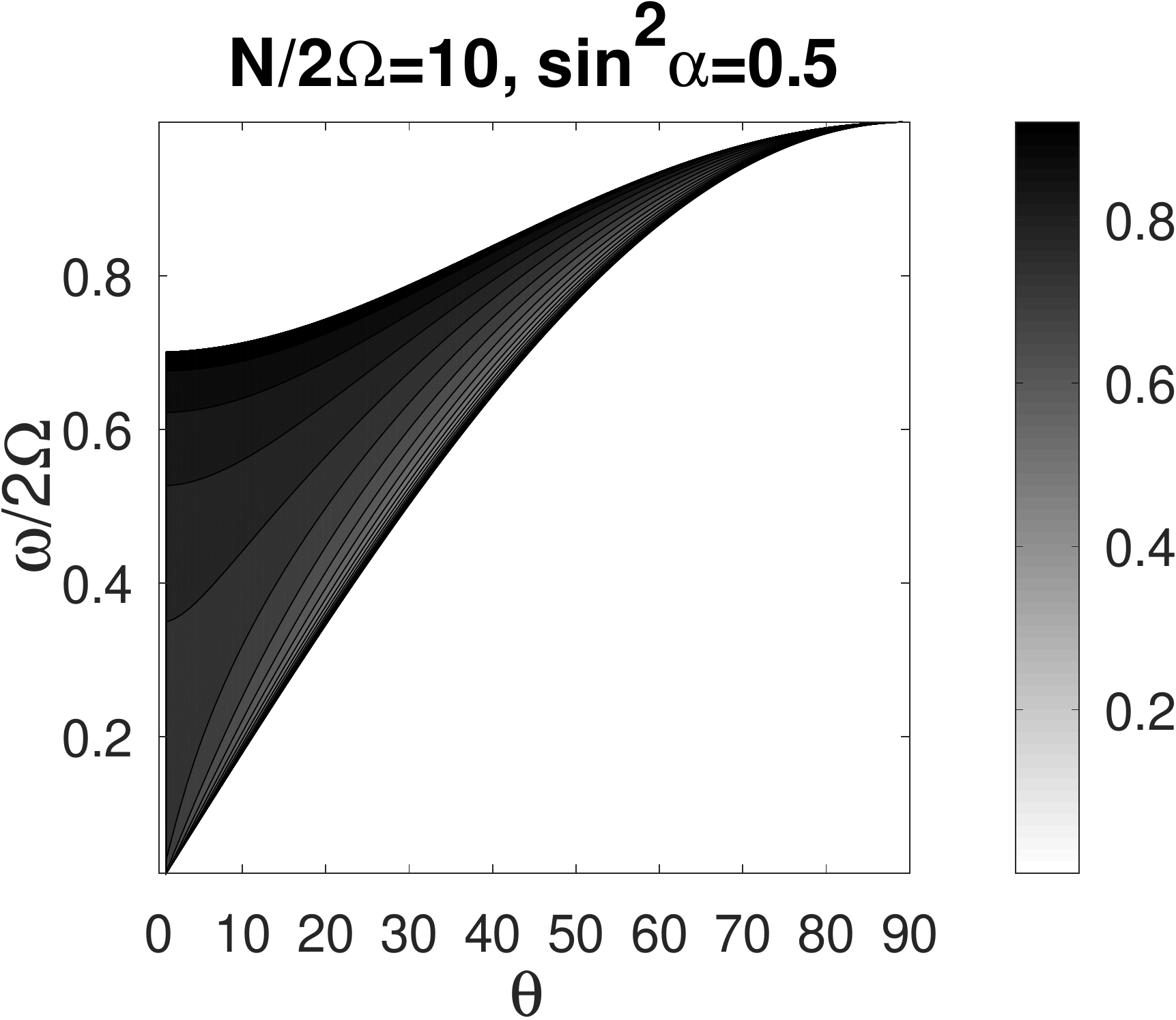}}
\subfigure[]{\includegraphics[scale=0.3]{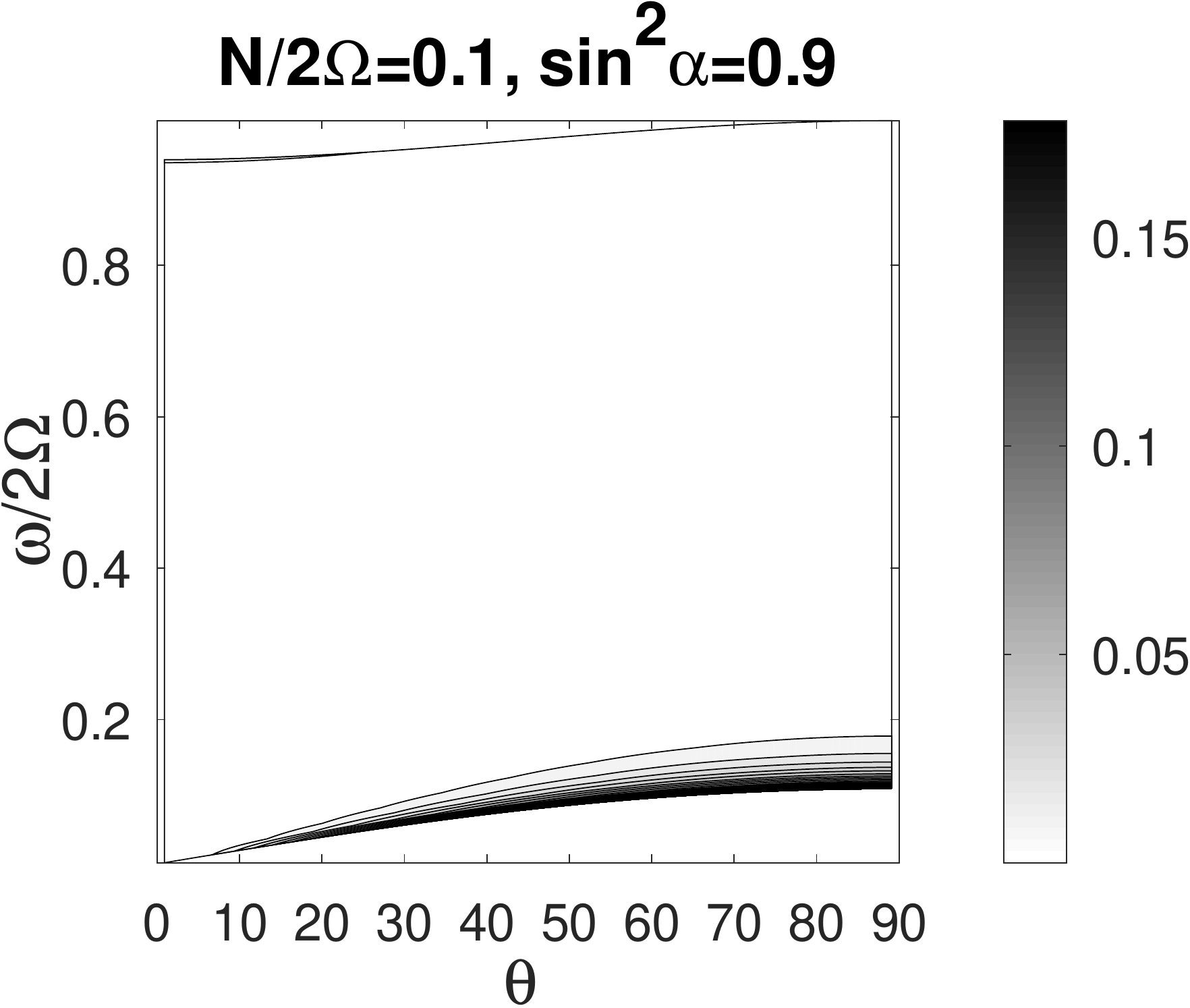}}
\subfigure[]{\includegraphics[scale=0.3]{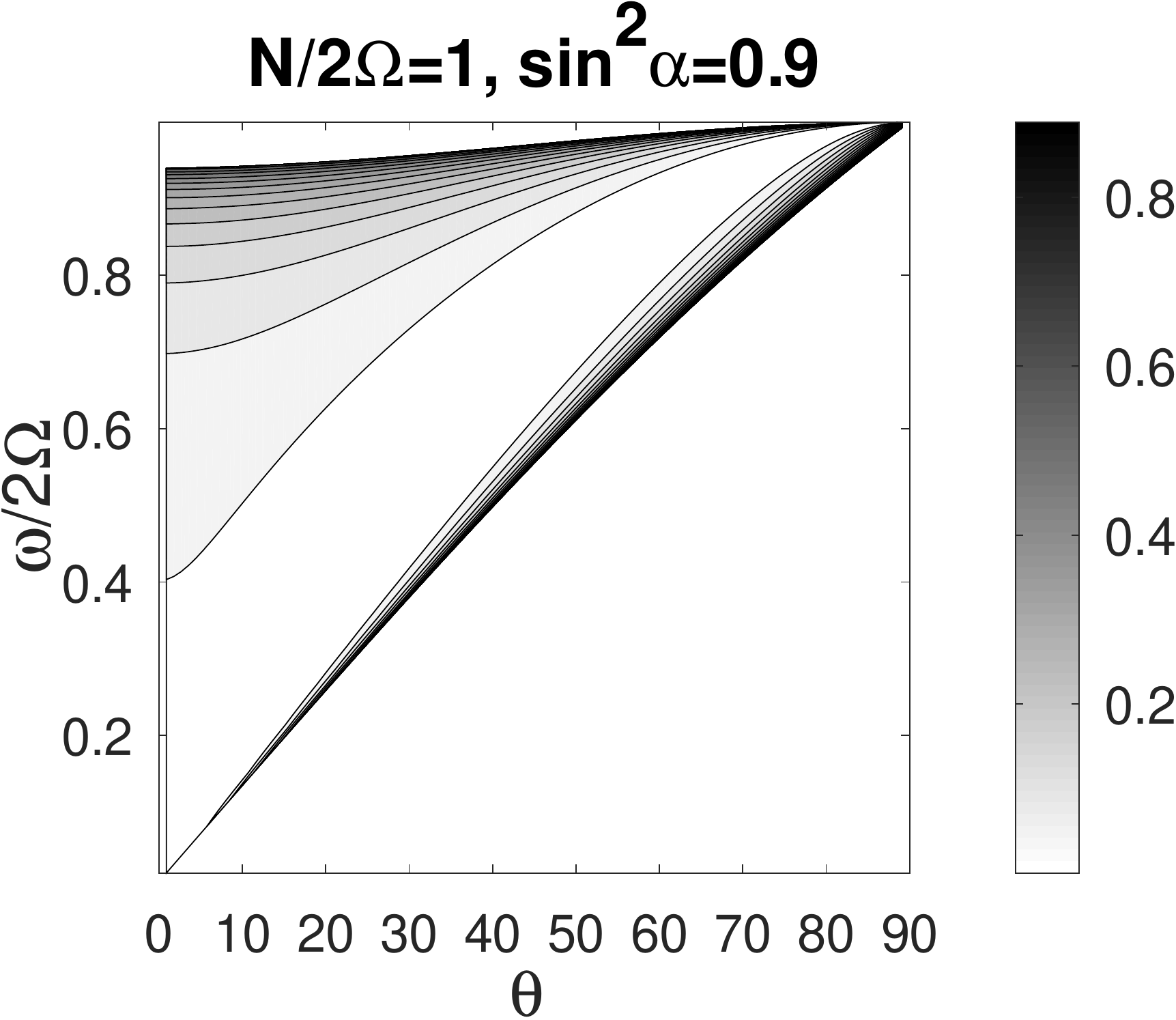}}
\subfigure[]{\includegraphics[scale=0.3]{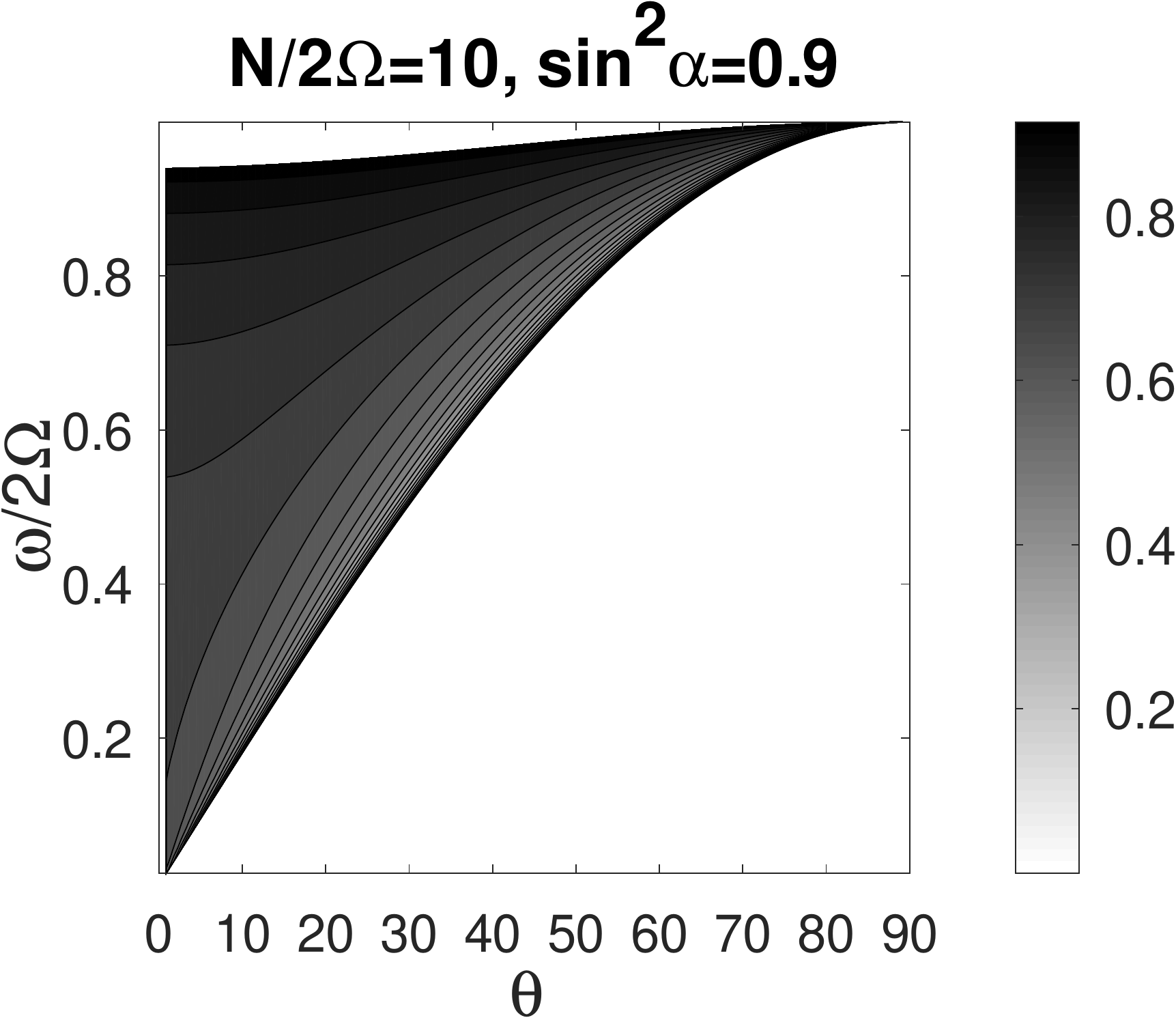}}
\caption{Contours of the reflection ratio $\eta$ versus wave frequency $\omega/2\Omega$ and latitude $\theta$ at different stratification strength $N/2\Omega$ and direction of wavevector $\sin^2\alpha$. The greyscale corresponds to the value of the reflection ratio as indicated by the bars on the right.}\label{fig2}
\end{figure}

In summary, in the original paper I used an inappropriate physical quantity, i.e. kinetic energy $\langle u'^2/2\rangle$, to measure the reflection and transmission ratios, and in the Erratum I use the vertical energy flux $\langle p'u_z'\rangle$ to measure the ratios to guarantee the sum of the two ratios to be 1. However, the major result still holds, i.e. rotation favors the wave transmission whereas stratification inhibits the wave transmission.

\end{document}